\documentclass{JHEP3} 
\pdfoutput=1
\usepackage{subfigure}
\usepackage{epsfig}
\usepackage{graphicx}
\usepackage{enumerate}	  
\usepackage{amsmath}
\usepackage{amsfonts}

\newcommand{\be}{\begin{equation}}	\newcommand{\bea}{\begin{eqnarray}}
\newcommand{\eea}{\end{eqnarray}} \newcommand{\ee}{\end{equation}}

\def\N{{\cal N}}		\def\L{{\cal L}}		\def\R{{\cal R}}		\def\J{{\cal J}}	\def\q{{\mathfrak q}}
\DeclareMathOperator{\tr}{Tr}			\def\l{\ell}			\def\G{{\cal G}}	\def\j{{\mathfrak j}}
				\def\H{{\cal H}}		\def\F{{\cal F}} 	\def\V{{\mathsf V}}
\def\n{{\mathfrak n}}			 \def\d{{\mathsf d}}
\def\O{{\cal O}}				\def\S{{\mathsf S}}

\title{Three-Charge Supertubes in a Rotating Black Hole Background} 
\author{Tehani K. Finch\\ Department of Physics and Astronomy, Howard University\\ Washington DC 20059, USA \\ 
\rm {email: tkfinch (at) howard.edu, lam137 (at) hotmail.com}}   

\abstract{The low velocity scattering of a D0-F1 supertube in the background of a BMPV black hole has been investigated in the moduli space approximation by Marolf and Virmani.  Here we extend the analysis to the case of the D0-D4-F1 supertube of Bena and Kraus.  We find that, similarly to the two-charge case, there is a critical value of the supertube circumferential angular momentum; above this value an adiabatic merger with the black hole cannot occur. By reconsidering the calculation of supertube angular momentum in the transverse direction, correspondence between the worldvolume and supergravity descriptions is established.  We also examine dynamical mergers and discuss their implications.} 


\begin{document}
\baselineskip=17pt    
\noindent\date{April 2009}
\section{Introduction}
In their original worldvolume formulation, supertubes are solutions of the Dirac-Born-Infeld (DBI) low-energy effective action for D-branes in Type IIA string theory \cite{mateos}.  They take the form of tubular D-brane configurations, possessing static electric and magnetic fields on the D-brane worldvolume that produce a nontrivial amount of angular momentum.  They also preserve the same supersymmetries as the three-charge rotating supersymmetric black hole, called the BMPV solution after the authors of \cite{bmpv}, which makes them ideal probes of such a black hole. 

Not being exact supergravity solutions, these `worldvolume supertubes' do not incorporate backreaction of the supertube on the geometry, and do not account for certain interactions.  However, researchers have also found `supergravity supertubes', which are exact solutions of Type IIA and Type IIB supergravity actions \cite{emparan,townsend3}. 
Recent work, $e.g.$ \cite{elvang, warner2, kraus3, wang3, elvang2}, has shown that supergravity supertubes are part of a larger class of solutions of supergravity, called black supertubes, that saturate a Bogomol'nyi-Prasad-Sommerfeld (BPS) bound.  
These supersymmetric solutions in general possess three charges, three dipole moments, two independent angular momenta and an event horizon.  Their reduction to five dimensions includes BPS black rings and BMPV black holes.  This family of solutions also contains BPS objects without event horizons, and the supertubes to which we refer in the present work belong to this family.  
Moreover, certain (non-supersymmetric) higher dimensional lifts of black rings, called black tubes, have been shown to correspond to excited states of two-charge and three-charge supertubes \cite{elvang, elvang2}.    

Three-charge worldvolume supertubes were introduced in \cite{bena}, with a contrasting realization being given in \cite{elvang}.  The agreement between the worldvolume and supergravity formulations is not exact.  As mentioned, generic three-charge supergravity supertubes possess three dipole moments and two angular momenta.    
In contrast, both of the worldvolume supertube descriptions provided in \cite{bena} have only two dipole moments and one nonzero angular momentum; that in \cite{elvang}, based on M-theory, has three dipole moments, but again only a single nonzero angular momentum.  The discrepancies are due to delocalized `flux term' contributions to the asymptotic charge and transverse angular momentum; these are beyond the scope of a worldvolume analysis.  
We will address these discrepancies in Section \ref{comparison}, in the process illuminating a gauge ambiguity in the worldvolume description. 

Despite its shortcomings, a great advantage of a D-brane worldvolume description 
over a full supergravity solution is that a time-dependent, low velocity scattering calculation can be performed straightforwardly.  Thus we can consider not only mergers of the supertube with a BMPV black hole that occur in the adiabatic limit, but also scattering behavior that occurs when the supertube is moving at a small velocity $(v\ll 1)$ with respect to the black hole.  The use of the DBI action to describe the physics of D-brane probes near black holes goes back more than a decade, one early example being \cite{strominger}.  More recently, the authors of \cite{virmani} undertook an investigation of D2-brane supertube scattering in the vicinity of a BMPV black hole.  The present treatment extends this analysis to the D6-brane worldvolume supertube of \cite{bena}, focusing on mergers of the supertube with the black hole.    

The results we find are similar to the two-charge case, although the expressions are more complicated.  It is found that adiabatic mergers can take place when the circumferential angular momentum $\j_1$ does not exceed a certain critical value $\j_{crit}$.   We establish a correspondence between the descriptions of adiabatic mergers for the worldvolume and supergravity supertubes.  For dynamical mergers, there exists, for certain ranges of the angular momenta, a stable equilibrium position of the supertube.  This configuration precisely matches the location and constraints on angular momenta of the corresponding stationary BPS solution.  

Furthermore, certain dynamical mergers, which we call fragmentation mergers, naively appear to decrease the black hole entropy.  All fragmentation mergers we consider are accompanied by a barrier in the effective potential.  This group of mergers includes, but is not limited to, overspinning mergers that violate the BMPV angular momentum bound $J^2<N_{D0} N_{D4} N_{F1}$.  If fragmentation mergers are allowed, they would presumably trigger a thermodynamic instability that would cause the black hole to fission.

In Section \ref{background} we give background information and definitions.  In Section \ref{sect3}  we discuss attributes and construction of the D0-D4-F1 worldvolume supertube of Bena and Kraus, and its adiabatic mergers with the black hole.  Section \ref{comparison} compares adiabatic mergers of the worldvolume and supergravity supertubes.  Section \ref{nonbps} discusses the range of validity of the moduli space approximation and treats dynamical scattering, $i.e.$ the non-BPS case of a slow velocity merger.  Section \ref{overspin} examines conditions for fragmentation and overspin of the black hole, and we present some brief concluding remarks in Section \ref{discussion}.  

\section{The BMPV Background} \label{background}
In the ten dimensional type IIA picture the black hole/supertube system has a D0-D4-F1 composition.  The full Type IIB supergravity solution for the BMPV metric and other background fields was obtained in \cite{herdeiro}; here we describe its IIA counterpart (related by a T-duality transformation on the $z$ direction).  It should be noted that the conventions here are related to those of \cite{bena} and much of the previous literature by the replacement  $\phi_2\rightarrow -\phi_2$.  The D4-branes are wrapped on the compact $T^4$, which has volume $V_{T^4}=(2\pi\ell)^4$; the F1 strings are wrapped on the compact $z$ direction, the length of $S^1_z$ being $2\pi R_z$.  
The type IIA supergravity solution for the BMPV black hole, with the metric expressed in the string frame, is  
\bea\label{back1}
ds^2 &=& -H_{D0}^{-1/2}H_{D4}^{-1/2}H_{F1}^{-1}(dt+\gamma_1(\theta) d\phi_1+\gamma_2 (\theta) d\phi_2)^2 + H_{D0}^{1/2}H_{D4}^{1/2}H_{F1}^{-1}dz^2\\ 
 & & + H_{D0}^{1/2}H_{D4}^{1/2}(dr^2 + r^2 d\theta^2 + r^2\sin^2\theta d\phi_1^2 +
r^2\cos^2\theta d\phi_2^2) + H_{D0}^{1/2}H_{D4}^{-1/2}ds^2_{T^4},\nonumber \\
e^{2\Phi} & = & H_{D0}^{3/2}H_{D4}^{-1/2}H_{F1}^{-1}, \label{back2} \\
C^{(1)} & = & (H_{D0}^{-1}-1)dt + H_{D0}^{-1}(\gamma_1 d\phi_1+\gamma_2 d\phi_2),\label{back3} \\
C^{(3)} & = & -(H_{D4}-1)r^2\cos^2\theta\, dz\wedge d\phi_1\wedge d\phi_2 -H_{F1}^{-1}dt\wedge
	dz\wedge(\gamma_1 d\phi_1+\gamma_2 d\phi_2), \label{c3}\\
B^{(2)} & = & (H_{F1}^{-1}-1)dt\wedge dz - H_{F1}^{-1}dz\wedge(\gamma_1 d\phi_1+\gamma_2 d\phi_2),\label{back5}
\eea
where $\Phi$ is the dilaton, $B^{(2)}$ is the NS-NS two-form, and $C^{(1)}$ and $C^{(3)}$ are the Ramond-Ramond (R-R) fields.  The noncompact space we parameterize using $\{r,\theta,\phi_1,\phi_2 \}$ with the origin $r=0$ at the horizon of the black hole.  The angles satisfy $0\le\phi_1,\phi_2<2\pi$ and $0\le\theta\le\frac{\pi}{2}$;  additionally they can be related to Cartesian coordinates $\{x^1,x^2,x^3,x^4\}$ through
\be x^1+ix^2=r\sin\theta e^{i\phi_1}; \quad\quad  x^3+ix^4=r\cos\theta e^{i\phi_2}. \nonumber  
\ee
We will also make use of the coordinates $\rho_1=r \sin\theta, \rho_2=r\cos\theta$.  Meanwhile, the toroidal directions are $\{x^6, x^7, x^8, x^9\}$ so that \be ds^2_{T^4} = (dx^6)^2+ (dx^7)^2 + (dx^8)^2+ (dx^9)^2. \ee

The angular momentum parameters $\gamma_1, \gamma_2$ and the harmonic functions $H_i$ are given by   
\bea
\gamma_1 &=& \frac{\omega}{r^2}\sin^2\theta, \quad \gamma_2=\frac{\omega}{r^2}\cos^2\theta, \\ 
H_{D0} &=& 1 +\frac{Q_{D0}}{r^2}, \quad H_{D4} = 1 +
\frac{Q_{D4}}{r^2}, \quad H_{F1} = 1 +\frac{Q_{F1}}{r^2}.
\eea
The quantities $Q_{D0},Q_{D4}, $ and $Q_{F1}$ are the charge parameters of the black hole (taken to be positive) 
with dimensions of $(\mbox{length})^2$, related to the integer numbers of D-branes $N_{D0}, N_{D4},$ and $N_{F1}$ by 
\be\label{ncharges}
N_{D0} = \frac{\ell^4 R_z}{g_s\, \alpha'^{7/2}} Q_{D0} , \quad N_{D4}  = \frac{ R_z}{g_s\,\alpha'^{3/2}}Q_{D4}, \quad N_{F1}  = \frac{\ell^4 }{g_s^2\, \alpha'^3}Q_{F1},
\ee
where $g_s$ is the type IIA closed string coupling constant \cite{malda}.  We restrict ourselves to situations for which $g_s\ll 1$, $g_s N_I\gg 1$. 
The BMPV black hole is characterized by angular momenta of equal magnitude in the planes of $\phi_1$ and $\phi_2$; the case we will consider is that of
\be\label{bhspin}
J_{1}  =  + J_{2}= \frac{\pi}{4 G_5} \omega\equiv J>0,
\ee
where $G_5$ is the gravitational constant in five dimensions.  This black hole has angular momenta in the $+\phi_1$ and $+\phi_2$ directions, and $J$ takes half-integer values since we set $\hbar=1$.  Furthermore, $J$ and $\omega$ are bounded; a regular horizon requires
\be\label{kerr}
J^2 < N_{D0} N_{D4} N_{F1} \quad\Leftrightarrow \quad\omega^2 <Q_{D0}\, Q_{F1}\, Q_{D4}. 
\ee
A violation of this bound would signify the presence of naked closed timelike curves (CTCs). 

The field strengths and Bianchi identity are 
\bea
&& \H^{(3)}=dB^{(2)},\quad\quad \G^{(2)}=dC^{(1)},
\quad\quad \G^{(4)}=dC^{(3)}+\H^{(3)}\wedge C^{(1)},\\
&& d\G^{(4)}+ \H^{(3)}\wedge \G^{(2)}=0.
\eea
Using $C^{(3)}$ and $C^{(1)}$ it is possible to introduce the ``magnetic" potentials $C^{(5)}$ and $C^{(7)}$, and these are necessary in our analysis.  They have the following field strengths and Bianchi identities \cite{ortin}:\footnote{Reference \cite{ortin} uses $(+,-,-,...,-)$ signature while we use $(-,+,+,...,+)$.  Therefore the relative signs of the terms in the Bianchi identities and the conventions for the dual fields in \cite{ortin} are the opposite of ours.}
\bea
&& -*\G^{(4)}= \G^{(6)} =dC^{(5)}, \quad \quad 
 *\G^{(2)}= \G^{(8)} =dC^{(7)} + \H^{(3)}\wedge C^{(5)} \\
&& d\G^{(6)}=0, \quad\quad 
d\G^{(8)}+ \H^{(3)}\wedge \G^{(6)}=0.
\eea
Notice that the field strength $\G^{(6)}$ is actually the $negative$ of $*\G^{(4)}$.  The magnetic potentials are found to be
\bea 
&&C^{(5)} = \Big( (H_{D4}^{-1}-1)dt + H_{D4}^{-1}(\gamma_1 d\phi_1 + \gamma_2 d\phi_2)\Big)\, 
\wedge dT^4\\ 
&&C^{(7)} = \Big(-(H_{D0}-1)r^2 \cos^2\theta\, dz \wedge d\phi_1\wedge d\phi_2 - H_{F1}^{-1}dt\wedge dz \wedge(\gamma_1 d\phi_1+\gamma_2  d\phi_2) \Big)\nonumber \\
 &&\wedge dT^4, \label{c7}
\eea
where $dT^4=dx^6\wedge dx^7 \wedge dx^8 \wedge dx^9$. 

What should be emphasized about the $dz\wedge d\phi_1\wedge d\phi_2$ term in $C^{(3)}$ and the $dz\wedge d\phi_1\wedge d\phi_2\wedge dT^4$ term of $C^{(7)}$ is that there is gauge freedom that allows replacement of $\cos^2\theta$ with, for example, $(-\sin^2\theta)$ in $C^{(3)}$ and $C^{(7)}$.  For a classical treatment (our chief concern here), solutions related by such a gauge transformation are physically equivalent.  However, the Wess-Zumino term of the Lagrangian does not automatically treat them as being equivalent, so ambiguities can arise from these gauge-dependent quantities.  We will appeal to exact supergravity results of Section \ref{comparison} to resolve the ambiguities.  It turns out that for a supertube with circumference in the $\phi_1$ direction, the replacement of $\cos^2\theta$ with $(-\sin^2\theta)$ is indeed required, so that the appropriate forms of $C^{(3)}$ and $C^{(7)}$ are given by
\bea  
&&C^{(3)}  =  (H_{D4}-1)\, r^2\sin^2\theta\, dz\wedge d\phi_1\wedge d\phi_2 -H_{F1}^{-1}dt\wedge
	dz\wedge(\gamma_1 d\phi_1+\gamma_2 d\phi_2), \label{c3sin}\\
&&C^{(7)} = \Big((H_{D0}-1)\, r^2 \sin^2\theta\, dz \wedge d\phi_1\wedge d\phi_2 - H_{F1}^{-1}dt\wedge dz \wedge(\gamma_1 d\phi_1+\gamma_2  d\phi_2) \Big)\nonumber \\
 &&\wedge\, dT^4 \, , \label{c7sin}
\eea
contrasting with the expressions of \cite{bena,virmani,bena09}.  This ensures accurate computation of the transverse angular momentum.\footnote{On the other hand, for a supertube with circumference in the $\phi_2$ direction, it is the original expressions (\ref{c3}) and (\ref{c7}) that give the accurate results.  The fact that a particular choice of $C^{(3)}$ and $C^{(7)}$ will not typically lead to predictions of symmetrical behavior between a $\phi_1$-oriented tube and a $\phi_2$-oriented tube is an indication of the above ambiguity.} 


\section{Bena-Kraus Supertubes}\label{sect3} 
\subsection{Construction}
The D6-brane worldvolume supertube of \cite{bena} is formed from a D6-brane with four dimensions wrapped on $T^4$.  Another dimension of the supertube, which we parameterize using $\sigma$, wraps a curve $S^1_{\sigma}$ in the uncompactified spacetime and its remaining direction we take to be along the $z$ axis.  Thus the worldvolume coordinates are $x^a=\{t,z,\sigma,x^6,x^7,x^8,x^9\}$.  The D6-brane possesses a gauge field on its worldvolume, $\tilde{F}$; for convenience we will work with the quantity $F =2\pi\alpha' \tilde{F}$, which expands as $F =\frac {1}{2}F_{ab}\, dx^a \wedge dx^b$.  Its general form will be 
\be
F = F_{tz} \,dt \wedge dz + F_{z \sigma }\, dz \wedge d\sigma + F_{t \sigma} \,dt \wedge d \sigma + 
F_{67} \,dx^6\wedge dx^7+F_{89} \, dx^8\wedge dx^9 .
\ee
In our conventions, $F_{t\sigma}$ and $F_{z \sigma}$ have dimensions of length and the other $F_{ab}$ are dimensionless.  We will use $V_6, V_{T^4},$ and $V_2$ to symbolize the spatial six-volume $(2\pi)^6 R_z \l^4$, compact four-volume $(2\pi\l)^4$ and two-volume $(2\pi)^2 R_z$ of the supertube, respectively.

The gauge field components can be interpreted as collections of superstrings and lower dimensional D-branes dissolved in the worldvolume of the D6-brane (see \it e.g. \rm \cite{douglas}).  Thus the supertube carries D0-brane charge $q_{D0}$, D4-brane charge $q_{D4}$, and fundamental superstring charge $q_{F1}$, in addition to D2 and D6-brane dipole moments $n_{D2}$ and $n_{D6}$,  often referred to in the literature as ``dipole charges" (see $e.g.$ \cite{elvang}).\footnote{The most general three-charge supertube also has an NS5-brane dipole moment, but that dipole is not captured by a proper worldvolume supertube.  It is possible to include an NS5 dipole charge in the ``superposition supertube" of \cite{bena}, which we discuss briefly in Appendix \ref{appa}.}  In principle there could also be D2-brane charge and a D4-brane dipole moment, but those will turn out to vanish for the construction utilized here.   

Our conventions for embedding the worldvolume into the spacetime 
($i.e.$ our gauge choice for the Lagrangian) are that we align the axes of the worldvolume coordinates $\{t, z, x^6, x^7, x^8, x^9\}$ with those of the spacetime ($i.e.$ the static gauge).  The position of the supertube in the noncompact space is labeled by $\{X^i\}=\{r,\theta,\phi_1,\phi_2\}$, and in the compact space by $\{X^p\}=\{z,x^6,x^7,x^8,x^9\}$.  Adopting the translational invariance of \cite{bena, virmani},  we require that $\{X^i\}$ depend on $\sigma$ and $t$, but not on $z$ or the $T^4$ directions, while $F_{ab}$ depends only on $t$.  Thus $X^i=X^i(t,\sigma)$.  We also require that the supertube has no motion or worldvolume dependence in the compact directions, $i.e.$ $X^p= const$.  
We will consider only the simplest circular embeddings; our primary focus will be on the choice $\phi_1=\sigma$.  In Section \ref{overspin} we shall also consider $\phi_2=\sigma$.  

In the worldvolume description supertubes can be discussed in terms of the DBI action,
\bea
\S &=& \int L \,dt = \int \L \, d^7 x = \int (\L_{BI}+\L_{WZ})\, d^7 x \label{action} 
= - \tau_{D6} \int d^7 x\, e^{-\Phi} \sqrt{-\det(g_{ab} +b_{ab} + F_{ab})} \cr
&+& \tau_{D6}\, \int \sum_{7-form\, terms} c^{(m)} \wedge e^{(F + b)^{(2)}}.
\eea
The lowercase variables ($g_{ab}$, $b_{ab}$, and $c^{(m)}$) refer to the pullbacks of the spacetime fields $G_{\mu \nu}$, $B_{\mu \nu}$, and $C^{(m)}$ to the D6 worldvolume.  The sum over $m$ in the Wess-Zumino (WZ) integral only includes terms in the wedge product that are 7-forms.  Choosing the opposite worldvolume parity to ours, \it e.g. \rm the ordering $\{t,\sigma,z,...\}$, would give an equivalent action for the supertube if the sign of the WZ term were also reversed.  
It should also be kept in mind that to arrive at the appropriate WZ term we must deal with the gauge ambiguity of Section \ref{background}.  The explicit Lagrangian appears in Section \ref{nonbps} and Appendix \ref{appa}.  

BPS configurations are realized in the supersymmetric limit, specified by
\be \label{bpslimit} 
F_{tz}=1, \quad F_{t\sigma}=0, \quad F_{67}=F_{89}, \quad \mbox{and} \quad \partial_t X^i=0.
\ee
This provides a symmetry under exchange of the pair $\{x^6, x^7\}$ with $\{x^8, x^9\}$.  Throughout the paper we will restrict ourselves to the case $F_{67}=F_{89}$.  The tensions of the Dp-branes take the form 
\be
\tau_{Dp}=\frac{1}{(2\pi)^p \,g_s\, \alpha'^{(p+1)/2}} \, , 
\ee
while the tension $\tau_{F1}$ of the F-strings is $\frac{1}{2\pi \alpha'}$.  

For the case of $F_{67}=F_{89}=B_0$, the supertube charges 
take the form \bea
q_{D4} & = & \frac{\tau_{D6}}{\tau_{D4}} \int dz \,d\sigma F_{z \sigma}=\frac{R_z}{\alpha'}F_{z \sigma},\label{charge4} \\
q_{D2} &=&\frac{\tau_{D6}}{\tau_{D2}} \int \, dz \, d\sigma\, (dx^6 dx^7 F_{z \sigma} F_{67} + dx^8 dx^9 F_{z \sigma} F_{89}) \label{q2eq} \\ 
&=& 2\frac{\l^2 R_z}{\alpha'^2}F_{z \sigma} B_0, \nonumber \\
 q_{D0} &=& \frac{\tau_{D6}}{\tau_{D0}} \int dT^4\, dz \, d\sigma\, F_{z \sigma}\,F_{67}\, F_{89}  =\frac{\l^4 R_z}{\alpha'^3}F_{z \sigma} B_0^2, \\
 q_{F1} & = & \frac{1}{\tau_{F1}}\int  dT^4\, d\sigma\frac{\partial {\cal L}}{\partial F_{tz}}.
\eea
These normalizations ensure that the charges take integer values.  Since 
\be
\frac{q_{D0}}{q_{D4}}=\frac{\l^4}{\alpha'^2}B_0^2, 
\ee
all expressions involving $B_0^2$ can also be written in terms of the ratio of supertube charges.  It is also significant that $q_{D2}$ satisfies \be\label{qd2} q_{D2}= 2\sqrt{q_{D0} q_{D4}}\, ,\ee 
showing that $q_{D2}$ is not an independent charge.

Our interest is in a supertube that will be T-dual to a D1-D5-P configuration, so we require it to carry D0, D4, and F1 charge, but no D2 charge.  Similarly we require it to possess D2 and D6 dipole charge but no D4 dipole charge.  Eq. (\ref{q2eq}) shows that a single D6-brane tube is inappropriate for such a task.  Consequently, we construct the supertube out of an even number $k$ of coincident D6-branes that are expected \cite{warner2, bena} to form a marginally bound state\footnote{Actually demonstrating that a such a state is in fact bound, $i.e.$ has a discrete energy spectrum, is nontrivial.  It has been achieved for the two-charge supertube in \cite{marolf2, ohta2}.}.  (This is done with the understanding that  $k$ is small enough for the DBI approximation to hold, which implies 
$g_s k \ll 1$.)  Half of these D-branes have $F_{67}=F_{89}=+B_0$; the others have $F_{67}=F_{89}=-B_0$, and thus have the opposite sign of $q_{D2}$, but are otherwise identical to the rest.\footnote{The construction of \cite{bena} is in flat spacetime and does $not$ require the branes to lie in the same plane or have the same size.  Here, however, we will eventually place the supertube near a black hole, and we want to ensure that the bound state will be maintained when the supertube has a small velocity.  Thus we require that the scattering behavior of all $k$ branes be identical.  This can be achieved if the branes all have the same embedding and physically differ by no more than the sign of $B_0$, since, as we will see, the dynamics depends on $B_0^2$ rather than $B_0$ itself.}  The $F_{ab}$ become diagonal $k\times k$ matrices $\F_{ab}$ and total charge is given by tracing over the  matrices.   Such a configuration has $\F_{67}= \F_{89}$; $\tr \F_{67}=\tr \F_{89}=0$; and $\F_{tz}$, $\F_{z \sigma}$, $\F_{67} \F_{89}$ all proportional to the unit matrix.  

The net D2 charge is eliminated because $\F_{z \sigma} \F_{67}$ and $\F_{z \sigma} \F_{89}$ have vanishing trace, and analysis of the open string spectrum indicates that there is no danger of a tachyon instability from the dissolved D2 and $\overline{D2}$-branes \cite{kraus2}.  
In contrast, the D0, D4, and F1 charges merely obtain a factor of $k$: 
\bea   
\q_{D4} &=& k\, q_{D4}, \quad \q_{D0}=k\, q_{D0}, \quad \q_{F1}=k\, q_{F1}, \\
\q_{D2}&=& \frac{R_z\l^2}{\alpha'^2}(\tr(\F_{z \sigma} \F_{67})+\tr(\F_{z \sigma} \F_{89})) = 0. 
\eea
For variables such as charge, angular momentum, and dipole charge, we use Fraktur letters $(\q,\j,\n)$ to denote quantities that describe the supertube as a whole, and italic type $(q,j,n)$ for those that correspond to one of the constituent branes.  
At times we will label the charges as $\{\q_I\}$ where $I=1,2,3$ and  $\{\q_1, \q_2, \q_3\}=\{\q_{D0},\q_{F1},\q_{D4}\}.$

The dipole charges $\n$, expressed in units in which they take integer values, take the form
\bea
\n_{D6}&=&\tr(I_k)=k, \label{nd6}\\ 
\quad \n_{D4}&=& \frac{\tau_{D6}}{\tau_{D4}}\frac{V_6}{(2\pi\l)^2\, V_2}(\tr \F_{67}+\tr \F_{89})  \\
&=& \frac{\l^2}{\alpha'}(\tr \F_{67}+\tr \F_{89})=0, \nonumber \\ 
\quad \n_{D2} &=& \frac{\tau_{D6}}{\tau_{D2}}\frac{V_6}{V_2}\tr(\F_{67} \F_{89}) = k  \frac{\l^4}{\alpha'^2}B_0^2, \label{nd2}\\
\quad \n_{NS5}&=& 0,
\eea
where $I_k$ is the unit $k\times k$ matrix.  The dipole $\n_{D4}$ vanishes for the same reason as $\q_{D2}$; both are proportional to $ (\tr \F_{67}+\tr \F_{89})$.  
The final dipole charge $\n_{NS5}$ is not captured in a classical D6-brane worldvolume treatment \cite{bena}, and is set to zero, leaving us with three nonzero charges and two dipole charges.  The case of a three-charge supertube with only one dipole charge, as well as that of two charges and two dipoles, is pathological and always contains CTCs \cite{bena4}.  In our case, for the  geometry near the supertube to be free of CTCs we need \cite{elvang, bena4}
\be\label{noctcs}
\frac{\n_{D2}}{\n_{D6}}=\frac{\q_{D0}}{\q_{D4}}\quad \rightarrow \quad \frac{\n_{D2}}{\n_{D6}}=\frac{q_{D0}}{q_{D4}}
= \frac{\l^4}{\alpha'^2}B_0^2,
\ee
which does indeed follow from the preceding equations.  

\subsection{Bena-Kraus Supertubes near a BMPV Black Hole}
The angular momenta in the $\phi_1$ and $\phi_2$ directions are $j_1$ (the circumferential angular momentum) and $j_2$ (the transverse angular momentum).  The circumferential angular momentum is essential for the stability of a supertube against collapse \cite{mateos}, and so $j_1\neq 0$; the focus here is on a supertube with $j_1>0$.  If $F_{z\sigma} >0$ the supertube charges have the same signs as those of the black hole (which were defined to be positive); the choice $F_{z\sigma}<0$ for the same $j_1$ would lead to supertube charges with signs opposite to the corresponding black hole charges, {\it i.e.} anti-branes.   As in \cite{bena} we take $F_{z \sigma}$ to be positive throughout the paper.  It should be noted that that there are actually $two$ distinct supertube configurations that possess a supersymmetric limit for the black hole background of (\ref{back1})-(\ref{back5}).  The one considered here has $F_{z\sigma}>0$ with $\j_1>0$ and (in BPS configurations) $\j_2>0$; like the black hole it carries angular momentum in the $+\phi_1$ and $+\phi_2$ directions.  There is another supertube for this background whose supersymmetric limit has $\{F_{z\sigma}<0, \j_1<0, \j_2<0\}$.\footnote{Additionally, there are two supertube configurations for the \it oppositely \rm charged black hole that has $C^{(1)}, C^{(3)}$ and $B^{(2)}$ given by the negative of those in (\ref{back3})-(\ref{back5}).  These configurations require $F_{tz}=-1$ in the supersymmetric limit.}  

Unlike the other charges, $q_{F1}$ has a value that in general depends on the black hole charges, as well as the position and velocity of the supertube, as we will see in Section \ref{nonbps}.  Its form in a BPS configuration is 
\be\label{bps1}
q_{F1}=  2\pi V_{T^4}\frac{\tau_{D6}}{\tau_{F1}}\frac{(H_{D0} + B_0^2 H_{D4} )\,r^2\sin^2\theta}{F_{z \sigma}} = \frac{\tau_{D6}V_6 (H_{D0} + B_0^2 H_{D4} )\,r^2\sin^2\theta}{q_{D4}},
\ee
where $V_6$ is the full spatial six-volume $(2\pi)^6 R_z \ell^4$.  

The angular momentum $\j_1$, obtained from the Lagrangian (\ref{fullarg6}) while letting $\phi_1$ momentarily depend on time, is 
\be\label{j1eq}
\j_1 = k\,j_1= \int dz\, d\sigma\, dT^4 \,\frac{\partial {\cal L}}{\partial \dot{\phi}_1}\Bigg{|}_{BPS}   = k \, q_{F1} q_{D4}.
\ee
The symbol $|_{BPS}$ denotes the BPS limit, specified by the conditions of (\ref{bpslimit}).  Note that $\j_1 = \q_{F1} \q_{D4}/k$, not $\q_{F1} \q_{D4}$.    
Meanwhile, the supersymmetric value of $\j_2$ derived from the Lagrangian based on (\ref{c3}) and (\ref{c7}) would be given by
\bea\label{bpsang} 
\j_2 = k\,j_2 &=&  \int dz \, d\sigma \, dT^4 \,\frac {\partial {\cal L}}{\partial \dot{\phi}_2}\Bigg{|}_{BPS}  = -\j_{crit}\cos^2 \theta,  
\quad \mbox{where} \\  \frac{\j_{crit}}{k} \equiv j_{crit} &=& \tau_{D6} V_6 (Q_{D0} + B_0^2 Q_{D4})= N_{D0} + B_0^2 \frac{\l^4}{\alpha'^2} N_{D4} \label{jcrit} \\ 
&  = & N_{D0}+ \frac{q_{D0}}{q_{D4}}N_{D4}. \nonumber
\eea
However, as mentioned at the end of Section \ref{background} and in Appendix \ref{appa}, the transverse angular momentum of the supertube arises from the gauge-dependent portion of the Lagrangian, $\L_{WZ}$.  Thus (\ref{bpsang}) cannot be trusted without additional input to single out the physically relevant gauge choice.  As we will see later, a full supergravity analysis reveals that the accurate result is 
\be \label{j2real} \j_2= \j_{crit}\sin^2 \theta, \ee 
implying that the appropriate versions of $C^{(3)}$ and $C^{(7)}$ are (\ref{c3sin}) and (\ref{c7sin}).  In the worldvolume formulation, $\j_2$ vanishes in the absence of the black hole (unlike the case of supergravity supertubes).  Parenthetically, we remark that the two-charge cases it happens that $j_{crit}=N_{D4}$ for the D0-F1 supertube,  and $j_{crit}=N_{D0}$ for the D4-F1 supertube. 

The supersymmetric value of the action is 
\be \label{actbps}
S=-\tau_{D6}\,k \int dt\, d\sigma\, dz\, dT^4 F_{z \sigma}\,(1+B_0^2)=-\int dt \, (V_{T^4}\, \tau_{D4} \q_{D4} + \tau_{D0} \q_{D0}).
\ee 
This leads us to the energy of the supertube, constructed from the Hamiltonian $\sum \tilde{p}_i\, \dot{\tilde{q}}_i- L$, giving 
\bea\label{en1}
E_{BPS} &=& \int dz\, d\sigma\, dT^4 \, \left( \frac{\partial \L}{\partial F_{tz}}F_{tz} - \L \right)\Bigg{|}_{BPS} = \tau_{D0} \q_{D0} +V_{T^4}\,\tau_{D4} \q_{D4} + 2\pi R_z \,\tau_{F1}\q_{F1} \cr 
&=& \frac{1}{g_s \sqrt{\alpha'}} 
\left(\q_{D0} + \frac{\l^4}{\alpha'^2}\q_{D4} + \frac{g_s R_z}{\sqrt{\alpha'}}\q_{F1} \right).
\eea  
$E_{BPS}$ is a minimum energy which saturates a BPS bound.  Now, the DBI action neglects any backreaction of the supertube on the spacetime metric and R-R fields, and consequently it will only be valid if the influence of the supertube on the background is negligible.  Thus in the paper we require
\bea 
g_s k\ll 1, \quad g_s N_I\gg 1, \quad |\j_1|, |\j_2| \ll J, 
&& \q_{D0} \ll N_{D0},\quad \q_{F1}\ll N_{F1},\quad \q_{D4}\ll N_{D4}, \nonumber \\ 
k^2\frac{\q_{D0}}{\q_{D4}}\ll \q_{F1}, && k\,\q_{D0}\ll\j_{crit}, \quad k\,\q_{D0}\ll\j_1,\label{condbps}
\eea
(the second set of conditions arising from supergravity results of Section \ref{comparison}), which will supplemented by other restrictions in the dynamical case.


\subsection{Adiabatic Mergers with the Black Hole}\label{embed}
\begin{figure} 
\begin{minipage}[t]{2.8in}
\includegraphics[width=2.5in]{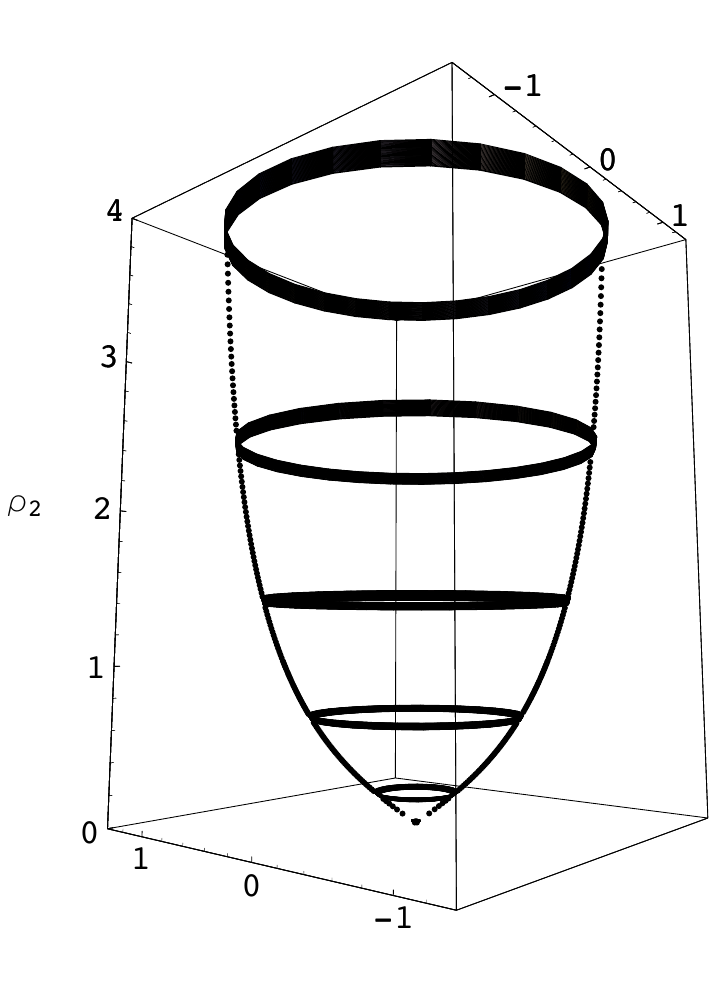}  
\renewcommand{\baselinestretch}{1}
\caption[A set of configurations of a BPS supertube for $j_1\le j_{crit}$.]{A set of possible BPS configurations (in cylindrical coordinates $\{\rho_1,\phi_1,\rho_2\}$ where $\rho_2$ is on the vertical axis) of a supertube for a given value of $j_1$ when $j_1\le j_{crit}$.  Here $j_1/j_{crit}=\frac{2}{3}$.  The black hole horizon is at the origin $(\rho_1=0, \rho_2=0)$.  As the supertube gets closer to the origin, it becomes small enough to ``fit" inside the black hole, and thus adiabatically merge with it.  The supertube appears as a ring because the $z$ direction is suppressed.}    
\label{tubein}  
\end{minipage}
\hfill
\begin{minipage}[t]{2.8in}    
\includegraphics[width=2.5in]{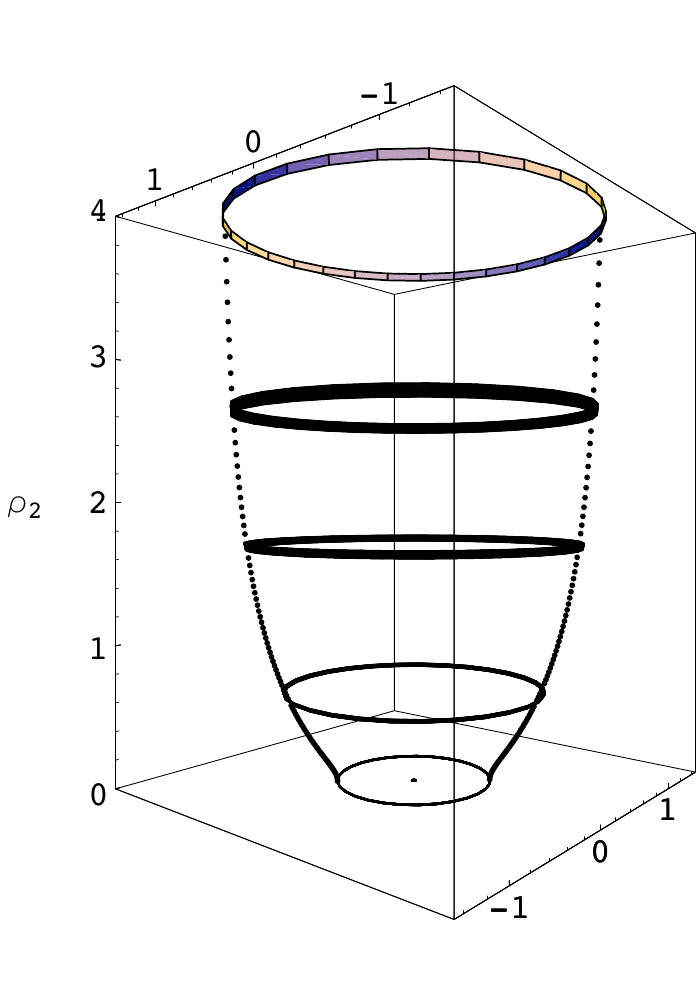}  
\renewcommand{\baselinestretch}{1}
\caption[A set of configurations of a BPS supertube for $j_1> j_{crit}$.]{A set of possible BPS configurations when $j_1> j_{crit}$.  Here $j_1/j_{crit}=1.18$.  As the supertube nears the black hole, its radius $R$ approaches the limiting value $R_{min}$,  
and the supertube cannot ``fit" inside the black hole (indicated by the dot at the center).  In both figures, the units are on the order of $10^{14}\sqrt{\alpha'}$, and the dots along the edges are included to indicate other intermediate locations.}
\label{tubeout}  
\end{minipage}
\end{figure}

Our choice of embedding exhausts the reparameterization freedom and thus represents a physical  orientation of the supertube with respect to the black hole.  Recalling the coordinates \be \rho_1=r\sin\theta=\sqrt{(x^1)^2+(x^2)^2}\, , \quad \quad\rho_2=r\cos\theta=\sqrt{(x^3)^2+(x^4)^2}\, ,\ee  
the (circular) cross section of the supertube lies in the $\rho_1$-$\phi_1$ plane; the line between the center of this circle and the black hole is just the $\rho_2$ axis, perpendicular to this plane.  Throughout the paper, the center of mass of the supertube moves only in the $\rho_2$-$\phi_2$ plane, so that this center either coincides with the black hole or remains ``directly above" it as shown in Figures \ref{tubein} and \ref{tubeout}.  The points of the supertube all have the same values of the coordinates $(r,\theta,\phi_2)$, (or equivalently $(\rho_1, \rho_2, \phi_2)$), which we will employ to specify its location.  
 
It turns out that there is a limited set of allowed locations for a supertube when the black hole is present.  Combining (\ref{charge4}) and (\ref{bps1}) gives\footnote{In the next two sections, many of the expressions are given in terms of $(q,j)$ rather than $(\q,\j)$.  This is merely for convenience and to make manifest the fact that these expressions are independent of $k$; they certainly apply to the entire supertube as well.}  
\be\label{bps_r}
j_1  = q_{F1} q_{D4}=\tau_{D6} V_6  (Q_{D0}+r^2 + B_0^2(Q_{D4}+r^2))\sin^2\theta \, .
\ee	
For a given $j_1$ and $j_2$, (\ref{bps_r}) and (\ref{j2real}) can be viewed as a restriction on the allowed combinations of $(r,\theta)$ for the BPS configurations.  Since we are required to have $j_1> 0$, (\ref{bps_r}) tells us that there are no BPS solutions of the DBI action for $\theta = 0$ at finite $r$, and that the allowed supersymmetric values of $r$ are given by 
\bea\label{r}
r^2 &=& \frac{j_1-j_{crit}\sin^2\theta}{\tau_{D6} V_6 \,(1+ B_0^2)\sin^2\theta}\,.   
\eea
Incorporating (\ref{j2real}) gives
\be\label{rbps}
r^2=\frac{1}{\tau_{D6} V_6 (1+B_0^2)}\frac{j_{crit}}{j_2}(j_1-j_2).\ee
BPS supertube locations must have $r>0$, as those precisely at $r=0$ would have null worldvolume \cite{virmani}.  Equations (\ref{j2real}) and (\ref{rbps}) then reveal more details related to the presence of the black hole:  BPS supertubes will satsify the relations 
\be \label{jcondbps} 0< j_2\le j_{crit}, \quad  j_1>j_2.\ee  
For future reference we add that the $\rho_1$ and $\rho_2$ values of the supertube satisfy
\be \label{rho1bps} \rho_{1\,(BPS)}^2 =\frac{j_1-j_2}{\tau_{D6} V_6 (1+B_0^2)},\quad\quad \rho_{2\,(BPS)}^2 =  \frac{j_1-j_2}{\tau_{D6} V_6 (1+B_0^2)} \frac{j_{crit}-j_2}{j_2} .\ee

The BPS configurations are stationary, and a given pair $\{\j_1,\j_2\}$ specifies $(r,\theta)$.  However, if only $\j_1$ is fixed, the supertube can in principle explore different regions of 
configuration space as long as its $r$ and $\theta$ coordinates satisfy (\ref{bps_r}).  During this process $\j_2$ varies, which implies that some external agent is applying a torque to the supertube.  The authors of \cite{bena} consider such a scenario, in which the supertube moves at infinitesimal velocity in the $r$ and $\theta$ directions while maintaining constant $\phi_2$ position.  
The energy changes only infinitesimally and $\{\j_1, \q_{D0},\q_{F1},\q_{D4}, \n_{D2}, \n_{D6}\}$ are conserved (even though dipole charges are not conserved in general).  The adiabatic limit of vanishing velocity is then invoked.  (In \cite{bena09} this analysis was extended to the case of a supertube merging with a black ring.)

Of particular interest are the BPS solutions in which the ring is arbitrarily close to the black hole horizon. The existence of these solutions gives rise to the idealized process of using the adiabatic limit to bring the supertube to the horizon and then infinitesimally farther, allowing it to fall into the black hole in an {\it adiabatic merger}.  
It can be seen from (\ref{j2real}) and (\ref{r}) that when the supertube reaches $r=0$, we have $j_1=j_2=j_{crit}\sin^2\theta$.  Hence, the angular momenta of the final product of the merger are still equal; it turns out the result is just another BMPV black hole.  Throughout the merger, the system is treated as though it were a BPS configuration.  The fact that there are in actuality no BPS supertube locations at the horizon itself further distances this merger from an actual physical process, however.  For actual physical motion, analysis of the dynamical moduli-space Lagrangian is needed.  This will be treated in Section \ref{nonbps}.

The embedding radius $R=R(\vec{X})$ of the supertube is given by (see Appendix \ref{appb})
\be\label{Rsq}
R^2 = H_{D0}^{1/2}H_{D4}^{1/2}r^2 \sin^2\theta=(Q_{D0}+r^2)^{1/2}(Q_{D4}+r^2)^{1/2}\sin^2\theta. 
\ee
In the asymptotically flat region at large $r$, $R^2$ approaches $r^2\sin^2\theta$, which in turn approaches a constant.  The relation (\ref{bps_r}) between allowed locations of the supertube and its physical quantities shows that this is just 
\be \label{Rinfty} R^2(r\rightarrow\infty)= \frac{j_1}{\tau_{D6} V_6 (1+ B_0^2)}\equiv R_{\infty}^2. 
\ee
This constant is independent of the black hole charges, as we might expect.  In fact, it is precisely the value obtained for a supertube in a Minkowski background.  We can also use (\ref{r}) and (\ref{rbps}) to obtain expressions that depend solely  on $\theta$, or solely on $r$, that are conducive to taking limits:
\bea\label{rsqtheta}
R^2&=&\frac{ \Big(j_1 + (N_{D0}-\frac{\l^4}{\alpha'^2} N_{D4})B_0^2\sin^2\theta\Big)^{1/2}\Big(j_1- (N_{D0}-\frac{\l^4}{\alpha'^2} N_{D4})\sin^2\theta\Big)^{1/2} } {\tau_{D6} V_6 (1+ B_0^2)}\, \\[2mm]
&=&  (Q_{D0}+r^2)^{1/2} (Q_{D4}+r^2)^{1/2}\frac{j_1}{\tau_{D6} V_6 (Q_{D0}+B_0^2\, Q_{D4} + (1+ B_0^2)\, r^2)}.\label{rsqr}
\eea  
Inspection of Equations (\ref{rsqtheta}) and (\ref{rsqr}) shows that $R^2$ reduces to $R_{\infty}^2$ for any $\theta$ or $r$ in the special case of $N_{D0}=\frac{\l^4}{\alpha'^2} N_{D4}$, \it i.e. \rm when $Q_{D0}=Q_{D4}$.  

\begin{figure} [t]     
\begin{minipage}[t]{2.8in}
\includegraphics[width=2.5in]{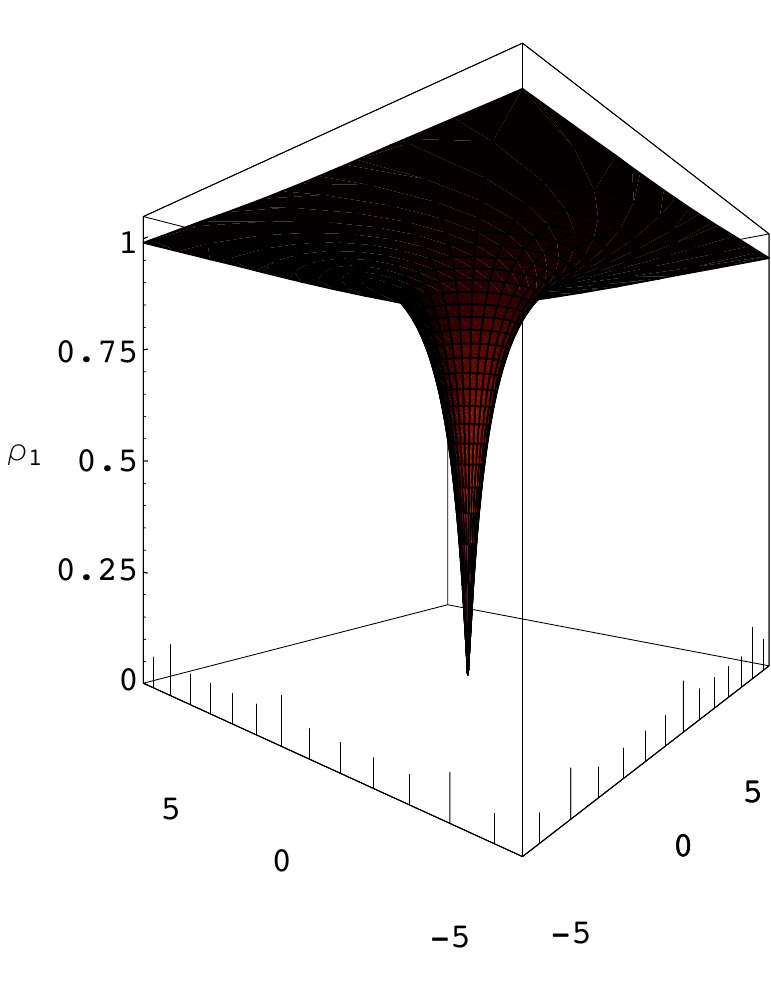}
\renewcommand{\baselinestretch}{1}
\caption[The configuration space of BPS locations for $j_1\le j_{crit}$.]{The configuration space (in cylindrical coordinates $\{\rho_2,\phi_2,\rho_1\}$ where $\rho_1$ is on the vertical axis) of BPS solutions for the location of a supertube for a given value of $j_1$ when $j_1\le j_{crit}$.  Solutions exist for all $\{\rho_1,\rho_2\}$ such that $0<\rho_1<1$ and $\rho_2>0$.  
Here $j_1/j_{crit}=\frac{2}{3}$.}    
\label{modin}  
\end{minipage}
\hfill
\begin{minipage}[t]{2.8in}
\includegraphics[width=2.5in]{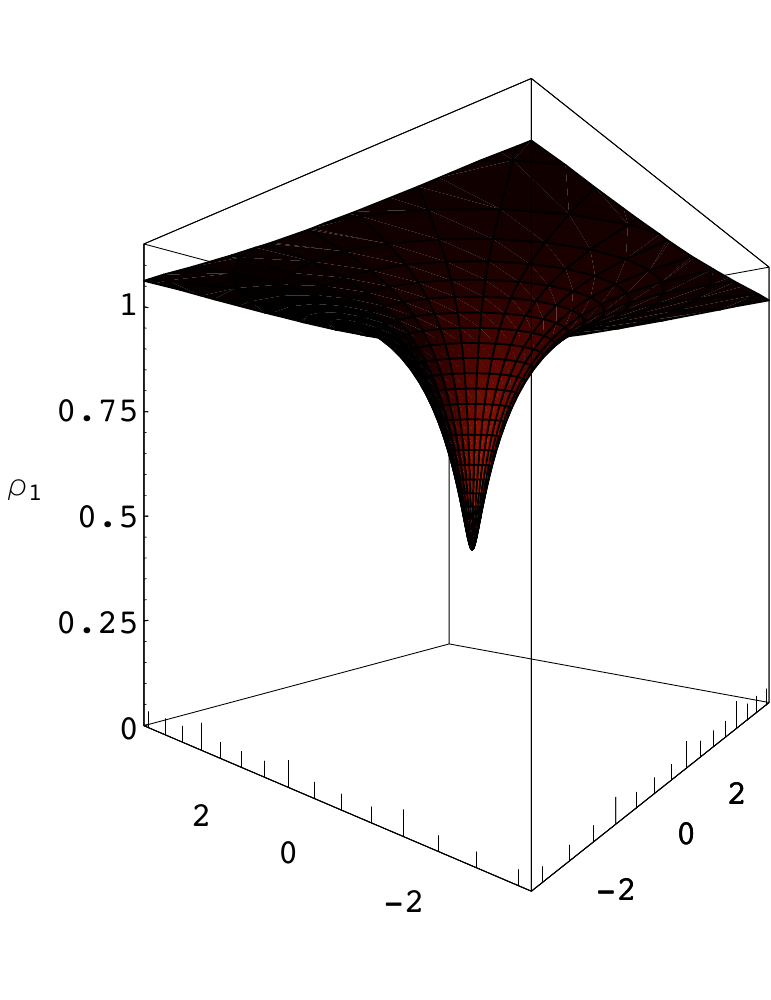}
\renewcommand{\baselinestretch}{1}
\caption[The configuration space of BPS locations for $j_1> j_{crit}$.]{The configuration space of BPS solutions when $j_1> j_{crit}$.   There are no BPS solutions for $\rho_1<0.41$.  Here $j_1/j_{crit}=1.18$.  In both figures, distances are in units of $10^{14}\sqrt{\alpha'}$.}
\label{modout}  
\end{minipage}
\end{figure}

Our supertubes are best visualized using the coordinates $\rho_1$ and $\rho_2$.  This facilitates clarification of the meaning of large values of $r$.  For our embedding, the limit $\rho_2\rightarrow\infty$ signifies the center of the supertube moving very far away from the black hole, though remaining directly above it.  A schematic of BPS configurations of a supertube, which appear ring-like due to suppression of the $z$ direction, appears in Figures \ref{tubein} and \ref{tubeout}.  In this work we only consider mergers in which $j_1$ is conserved.  Thus, (\ref{r}), which applies during adiabatic mergers, tells us that in such a process $\rho_1$ has a maximum value of $R_{\infty}$.  (We emphasize that this is not the case for the {\it dynamical} mergers discussed in Chapter \ref{nonbps}.)  This means that at large $r$ we have $r\approx \rho_2$ along with $\rho_1=R_{\infty}$, and the limit $r\rightarrow\infty$ refers to $\{\rho_1\rightarrow R_{\infty}, \rho_2\rightarrow\infty, \theta\rightarrow 0\}$.\footnote{Such a limit will be invoked again, when defining the so-called `Region III' of Section \ref{nonbps}.}  The configuration space, in which every point represents a supertube location, is shown in Figures \ref{modin} and \ref{modout}. 

Now consider small values of $r$.  Since $\sin^2\theta\leq 1$ we can see from (\ref{r}) that it is possible to adiabatically bring the supertube to the horizon at $r=0$ if  
\be \label{crown} j_1 \le   j_{crit}  = N_{D0} + B_0^2 \frac{\l^4}{\alpha'^2}N_{D4},
\ee 
as in Figure \ref{modin}.  
According to (\ref{jcrit}) and (\ref{Rsq}), as the supertube nears the horizon, its size decreases to
\be \label{fit1}
R^2(r=0)=\frac{j_1}{j_{crit}} Q_{D0}^{1/2} Q_{D4}^{1/2} \equiv R_0^2, 
\ee  	
while due to (\ref{r}), its $\theta$ value increases to \be \sin^2\theta_{merge}=j_1/j_{crit}. \ee 
The value of the vertical angle at the vertex of Figure \ref{tubein} is $2 \theta_{merge}$. 

On the other hand, if $j_1>j_{crit}$, then the supertube cannot reach $r=0$ (because it cannot reach $\rho_1=0$, as shown in Figure \ref{modout}).  Thus in $j_{crit}$ we find a limit to the circumferential angular momentum of a supertube that can merge with the black hole.  
We will see that this is limit also holds in a full supergravity analysis:  a supergravity supertube that undergoes an adiabatic merger with the black hole has an analogous bound on its circumferential angular momentum.

If $j_1 >   j_{crit}$, (\ref{r}) indicates that the closest the supertube can come is  
\be\label{rmin}
r^2=\frac{j_1-j_{crit}}{\tau_{D6} V_6 (1+ B_0^2)}\equiv r_{min}^2.
\ee  
This minimum distance occurs when $\sin\theta=1$, so using (\ref{Rsq}), 
the embedding radius at $r_{min}$ is found to be given by
\be\label{toobig1}
R_{min}^2= (Q_{D0}+r_{min}^2)^{1/2} (Q_{D4}+r_{min}^2)^{1/2} >R_0^2.  
\ee
In this case the black hole passes harmlessly through the center of the supertube, (e.g. the bottom ``ring" configuration in Figure \ref{tubeout}) while the embedding radius shrinks to its minimum value $R_{min}$.  The  
supertube then recedes, up the opposite side of the ``cone" in Figure \ref{modout}.  As it does so, it re-expands, with $R$ approaching $R_{\infty}$ as it gets farther away from the black hole.

Both the supertube and the black hole have descriptions in type IIB supergravity, which can be obtained from the IIA solutions by T-dualizing on the $z$ coordinate.  This procedure takes us from a D0-D4-F1 system to a D1-D5-P system, and in terms of the charge parameters $Q_{D1}$ and $Q_{D5}$ the embedding radius is now (Appendix \ref{appb})
\be R^2= (Q_{D1}+r^2)^{1/2}(Q_{D5}+r^2)^{1/2}\sin^2\theta, \ee
and we also have $Q_{D1}=Q_{D0}$ and $Q_{D5}=Q_{D4}$.  Comparison with (\ref{Rsq}) shows that this is the same embedding radius as before.  Near the horizon, the IIB BMPV metric 
looks like $AdS_3\times S^3\times T^4$.  

In \cite{kraus3} and \cite{bena} it was pointed out that  the $S^3$ has a radius (in the string frame) given by \be R_{S^3}^2=Q_{D1}^{1/2} Q_{D5}^{1/2}.  
\ee  
If the embedding radius of a BPS supertube at the horizon $R_0^2$ satisfies $R_0^2>Q_{D1}^{1/2} Q_{D5}^{1/2}$, it was argued that in the IIB picture the supertube will not ``fit" inside the near horizon region of the black hole, and thus that the supertube cannot adiabatically merge with it.  This reasoning holds here as well.  However, we are not aware of an extension of this argument to the case of a dynamical merger, in which the tube has no obvious counterpart to $R$.  Indeed, it turns out that a supertube that is naively ``too big" to fit, $i.e.$ having $j_1>j_{crit}$, $can$ be pushed into the black hole (taking us into the dynamical regime), as we will discuss in Section \ref{nonbps}.  Furthermore, our analysis in the dynamical case will extend to $r\rightarrow\infty$ only if the supertube moves along a path similar to that of an adiabatic merger.   


\section{Comparison of Bena-Kraus and Supergravity Supertubes}\label{comparison} 

The subject of supersymmetric black rings in the presence of a BMPV black hole was treated in \cite{warner2} and the issue of adiabatic mergers of the ring with the black hole appeared in \cite{bena2, bena3}.  The authors, using the same embedding as presented here, found an exact BPS supergravity solution for the eleven dimensional lift of the black ring + black hole system; this solution can also be expressed in terms of Type IIA or IIB supergravity.  These are solutions in which the black ring, like our supertube, lies directly above the black hole when $\theta<\frac{\pi}{2}$.  The specific case of $\theta=\frac{\pi}{2}$, in which the black ring and black hole lie in the same plane and are concentric, had been found in \cite{gauntlett1, gauntlett2}.  In the following section we will label the charges and dipole moments by those of the corresponding IIA solution.  

In this supergravity solution 
it is possible to set one of the dipole moments of the black ring to zero, but then the causal structure of the geometry must be checked.  It turns out that eliminating the closed timelike curves forces the result to be a BMPV black hole together with a ``black ring" of zero horizon area, \it i.e. \rm a supergravity supertube.  Thus our configuration is a three-charge, two-dipole supergravity supertube in a BMPV background.  Now, such a supertube has a naked curvature singularity at its core \cite{elvang, bena4}.  However, Reference \cite{elvang2} was able to describe this supertube, for a certain range of angular momentum, as the supersymmetric limit of a family of black rings that do have regular horizons. It was therefore suggested that despite the naked singularity, this solution should be regarded as physically sensible.  
Moreover, in \cite{bena4} the singularity was attributed to brane sources and described as not pathological.  This perspective will be adopted in the present work, from which we conclude that this singularity does not vitiate the analysis presented herein.  

The supergravity solution has long range R-R fluxes, and the angular momentum and charge measured by flux integrals at infinity (asymptotic charges) differ from those determined by flux integrals at the supertube itself \cite{kraus3}.  The latter quantities, called `microscopic,' are those suitable for comparison with the worldvolume supertube, and represent the number of D-branes and F-strings that comprise the black ring in its string theory formulation.\footnote{The microscopic charges are not conserved, and there has been some dispute about their precise physical significance.  See $e.g.$ \cite{horowitz, warner4}.}  Components of charge and angular momentum not localizable to the supertube are `flux terms.'  This section aims to establish correlations between the BPS worldvolume and supergravity supertubes in a BMPV background, along lines similar to those in Section 8 of \cite{elvang}.  The results of Section \ref{nobh} of the present paper were also presented there, although mostly in the context of the ``superposition supertube", which we briefly describe in Appendix \ref{appa}.

We orient the supergravity supertube in the same manner as the worldvolume supertube.  For ease of comparison with \cite{bena2} (with the caveat that our $\phi_2$ is the negative of theirs, as mentioned in Section \ref{background}, so that the corresponding $\phi_2$ angular momentum terms will differ by a sign), we adopt the following notation:  the symbols 
\be  \quad \{\zeta_1, \zeta_2, \zeta_3\} = \{\zeta_{D0}, \zeta_{F1}, \zeta_{D4}\} 
\quad\mbox{and}\quad
\{\d^1, \d^2, \d^3\} = \{\d_{D6}, \d_{NS5}, \d_{D2}\} 
\ee 
label the microscopic charges and dipole charges of the supergravity supertube, while
\bea 
\{N_1, N_2, N_3\} &=& \{N_{D0},N_{F1},N_{D4}\}, \quad\{\q_1, \q_2, \q_3\} =\{\q_{D0}, \q_{F1}, \q_{D4}\}, \quad \mbox{and} \\ \{\n^1, \n^2, \n^3\} &=& \{\n_{D6}, \n_{NS5}=0, \n_{D2}\} \nonumber
\eea
represent the black hole charges, worldvolume supertube charges and worldvolume supertube dipole charges respectively.  Since the worldvolume supertube necessarily has $\n_{NS5}=0$, we restrict the supergravity supertube to have $\d_{NS5}=0$.  All the $\{\zeta_I, \d^I, N_I, \n^I,  \q_I\}$ are integer-valued.  

\subsection{Isolated Supergravity Supertubes} \label{nobh}
Some context for later results is provided by the supergravity description of the supertube without a black hole present.  Contributions $\j_{\Delta}, \, \j_{\zeta}$ and $\j_c$ to the angular momenta of the supergravity solution take the forms \cite{elvang, bena2}
\bea\label{jdelt}
\j_{\Delta} &=& \tau_{D6} V_6\, r^2\sin^2\theta \, \left(\d_{D6}+\frac{\alpha'^2}{\l^4}\d_{D2} + \frac{\sqrt{\alpha'}}{g_s R_z} \d_{NS5}\right) \cr &=& \tau_{D6} V_6\, r^2\sin^2\theta \, \left(\d_{D6}+\frac{\alpha'^2}{\l^4}\d_{D2}\right), \\ 
\j_{\zeta} &=& \frac{1}{2} \d^{I} \zeta_{I}, \label{jzett}\\  \j_c &=& \frac{1}{6}C_{IJK} \d^I \d^J \d^K = 0,
\eea
where the microscopic angular momentum is $\j_{\Delta}$, and $C_{IJK}=|\epsilon_{IJK}|$.

The asymptotic charges of the spacetime we will denote by $\Xi_I$, and the asymptotic spacetime angular momenta by $\J_1$ and $\J_2$.  These are given by 
\bea\label{jy1}\J_1 &=& \j_{\Delta}+ \j_{\zeta}+ \j_c = \j_{\Delta}+ \j_{\zeta}, \\ \J_2 &=&  \j_{\zeta} + \j_c = \j_{\zeta},  \label{jy2} \\  \Xi_I &=& \zeta_I + \frac{1}{2} C_{IJK} \d^{J} \d^{K} \equiv \zeta_I  + \zeta_I^c. \label{cha62}\eea 
Unlike the worldvolume case, here $\J_2$ is nonzero even without a black hole.  The terms $\j_{\zeta}$, $\j_c$ and $\zeta_I^c$ are flux terms.  Of course, since there are only two nonzero dipole charges, $\j_c$ vanishes.

Meanwhile, using  (\ref{bps1}), (\ref{nd6}), (\ref{nd2}) and (\ref{j1eq}) we see that for the \emph{worldvolume} supertube,
\bea \label{j1wv}\j_1 &=& \frac{\q_{F1}\q_{D4}}{k}
=\tau_{D6} V_6\, r^2\sin^2\theta \, \left(\n_{D6}+\frac{\alpha'^2}{\l^4}\n_{D2}\right), \\ 
\j_2 &=& 0.\eea
The similarity of (\ref{jdelt}) and (\ref{j1wv}) superficially implies a match between $\j_{\Delta}$ and $\j_1$, which is valid in this instance.  However, \cite{elvang} presented arguments 
that in the general case of $\j_2\neq 0$ one should identify $\j_{\Delta}$ with $\j_1-\j_2$, and that approach will be adopted here.  The correspondence between worldvolume supertube quantities and their microscopic supergravity counterparts is denoted by the symbol ``$\Leftrightarrow$", so that 
\be \label{jdelt2}	\tau_{D6} V_6\, r^2\sin^2\theta \, \left(\n_{D6}+\frac{\alpha'^2}{\l^4}\n_{D2}\right) = (\j_1-\j_2)\Leftrightarrow \j_{\Delta} .\ee 
We now have
\be\label{identify}
\j_1-\j_2\Leftrightarrow \j_{\Delta}, \quad \n^I\Leftrightarrow \d^I, \ee 
along with \be \label{chargerep} \quad \q_I \Leftrightarrow \zeta_I. \ee
Using (\ref{jzett}) and (\ref{cha62}) this leads to  
\be \label{jzeta}
\{\zeta_1^c,\,\zeta_2^c,\,\zeta_3^c \}\Leftrightarrow \{0,\n_{D2}\,\n_{D6},0\} \quad\mbox{and} \quad \j_{\zeta}\Leftrightarrow \n_{D6}\,\q_{D0}=\n_{D2}\,\q_{D4} \ee 
after using (\ref{noctcs}).  

Now, the worldvolume and supergravity descriptions do not agree, because the worldvolume quantities do not contain the flux terms  $\j_{\zeta}$ and $\zeta_I^c$.  Let us introduce an augmented worldvolume charge, which is meant to correspond to the full supergravity spacetime charge, $i.e.$ 
\be \q_I^{(aug)}\Leftrightarrow \Xi_I. \ee
A key observation is that using the augmented worldvolume charge is all that is needed to reproduce the flux terms.  Thus the replacement
\be  \label{replace}\q_I\rightarrow \q_I^{(aug)}, \ee 
resolves the discrepancies between the two descriptions; the substitution (\ref{replace}) introduces effective charge and effective angular momentum that accounts for the flux terms.

Let us verify this claim.  After implementing (\ref{replace}), $\q_{D0}$ and $\q_{D4}$ remain unchanged, but for $\q_{F1}$, the worldvolume replacement and new supergravity correspondence are given by
 \bea 
 \label{replace2} \q_{F1}&\rightarrow& \q_{F1}+ \n_{D2}\,\n_{D6}, \\
 && \q_{F1} + \n_{D2}\,\n_{D6}\Leftrightarrow \zeta_2+ \zeta_2^c, \eea
with $\n_{D2}\, \n_{D6}$ acting as effective worldvolume charge.  Substitution of (\ref{replace2}) into (\ref{j1eq}) and using (\ref{nd6}) and (\ref{noctcs}) yields the accompanying change in $\j_1$:
\bea \j_1\rightarrow \frac{\q_{F1}\q_{D4}}{k} + \j_1^{\,(eff)} &=& \frac{\q_{F1}\q_{D4}}{k}
+ \n_{D2}\,\q_{D4},\\
&& \frac{\q_{F1}\q_{D4}}{k}+ \n_{D2}\,\q_{D4} \Leftrightarrow  \j_{\Delta}+ \j_{\zeta},
\eea
and since $\j_{\Delta}\Leftrightarrow \j_1-\j_2$ we also have a change in $\j_2$:
\bea \j_2 \rightarrow 0 + \j_2^{\,(eff)} &=& \n_{D2}\,\q_{D4}, \\
&& \n_{D2}\,\q_{D4} \Leftrightarrow \j_{\zeta}. \eea
It is seen from (\ref{jy1}), (\ref{identify}) and (\ref{jzeta}) that we can now make the correlations
\bea \label{totang1} \j_1+ \j_1^{\,(eff)} &\Leftrightarrow& \j_{\Delta}+ \j_{\zeta} =\J_1, \\  \j_2^{\,(eff)} &\Leftrightarrow& \j_{\zeta}=\J_2. \label{totang2} \eea 
Hence the substitution (\ref{replace}), (a corollary of the one discussed in \cite{elvang}), produces effective angular momenta $\j_1^{\,(eff)}$ and $\j_2^{\,(eff)}$, leading to worldvolume equivalents of $\J_1$ and $\J_2$.  However, our implementation of (\ref{replace}) is purely formal; a detailed description of its underlying physics remains elusive.  We note that the worldvolume counterparts of $(\j_{\Delta}, \zeta_I)\sim k$ while those of the flux terms $(\j_{\zeta}, \zeta_I^c)\sim k^2$.  The roots of the second set of conditions of (\ref{condbps}) are now apparent: they ensure that the flux term corrections to the worldvolume quantities are negligible. 

\subsection{Supergravity Supertubes Near a BMPV Black Hole}
We are now in a position to examine the changes that occur when in the vicinity of a black hole.\footnote{Note that what the authors of \cite{bena2} call the ``embedding radius" is what we call the coordinate $\rho_1=r\sin\theta$ of the supertube.  Also, what they call $\alpha$, for us is $\cot\theta$, so their $n^I N_I^{BH}/(1+\alpha^2)$ is our $\d^I N_I \sin^2\theta$.}  
One difference is that a term
\be \j_N = \d^I N_I \sin^2\theta 
\ee contributes to both angular momenta, and the microscopic angular momentum along the 
$\phi_1$ direction is now \be \j_T= \j_{\Delta} + \j_N. \ee 

As with the worldvolume supertube merger of Section \ref{sect3}, the merger of a supergravity supertube with the black hole requires an external torque in the $\phi_2$ direction, and thus $\J_2$ varies; meanwhile $\{N_I, \zeta_I, \d^I, \j_T,\J_1\}$ are held constant.  Specifically, $\j_{\Delta}$ and $\j_N$ change during the merger, while their sum $\j_T$ does not.  The final state of such a merger is merely another BMPV black hole, and it was shown in \cite{bena2} that this adiabatic merger, like the worldvolume one, involves no danger of lowering the entropy.

Other attributes of the supergravity solution are that for a merger to even take place,\be\label{crown2} \j_T \le \d^I N_I \ee must be satisfied, and during the merger we have  \bea\label{jay1}\J_1 &=& \j_T+ \j_{\zeta}+ J, \\ \J_2 &=&\j_N + \j_{\zeta}+  J,  \label{jay2} \\ \Xi_I &=& N_I + \zeta_I +\zeta_I^c ,\label{charge62}\eea   
where $J$ is again just the BMPV angular momentum. 

We now compare the supergravity and worldvolume quantities in the presence of the black hole.  Equations (\ref{nd6}) and (\ref{nd2}) show that for the worldvolume supertube
\be \n^I N_I =\j_{crit},  
\ee
allowing an identification of $\j_N$ with $\j_{crit}\sin^2\theta$.  Moreover, examining (\ref{bps_r}) and (\ref{jdelt}) confirms that we can equate $\j_1$ to the microscopic supergravity quantity $\j_T$.
Accordingly, we see that in the presence of the black hole (\ref{identify}) and (\ref{chargerep}) still hold, and we list them along with our other relations:
\bea
\j_1- \j_2 &\Leftrightarrow& \j_{\Delta}, \label{identify2}\\
\q_I &\Leftrightarrow& \zeta_I, \quad  \n^I\Leftrightarrow \d^I, \quad  \j_N \Leftrightarrow \j_{crit}\sin^2\theta, \quad \j_1\Leftrightarrow \j_T.  
\eea 
To give worldvolume counterparts to the full supergravity angular momenta, we again implement the replacement (\ref{replace2}), obtaining an effective angular momentum $\j_1^{\,(eff)}$ that adds to $\j_1$:   
\bea  \j_1^{\,(eff)} & = & \n_{D2}\,\q_{D4}, \nonumber \\
 \j_1& \rightarrow & \j_1+  \j_1^{\,(eff)} \nonumber \\
& = & \tau_{D6} V_6\, r^2\sin^2\theta \, \left(\n_{D6}+\frac{\alpha'^2}{\l^4}\n_{D2}\right) +\j_{crit}\sin^2\theta + \n_{D2}\,\q_{D4} \,. 
\eea

Now, we saw before that the worldvolume analysis does not give a gauge-invariant result for the transverse angular momentum $\j_2$ in the presence of a black hole.  Since we know the supergravity supertube is an exact solution of the supergravity equations of motion, it will guide our analysis.  Equations (\ref{jay2}) and (\ref{identify2}) tell us that the worldvolume $\j_2$ of the supertube should just be the quantity corresponding to $\j_N$.  Therefore, we \it demand \rm that the gauge-invariant $\j_2$ for the worldvolume supertube be 
\be \j_2= \j_{crit}\sin^2\theta. \ee 
This we assert is the proper supersymmetric value of $\j_2$ for our embedding, which validates the use of (\ref{c3sin}) and (\ref{c7sin}) for $C^{(3)}$ and $C^{(7)}$.  Equation (\ref{identify2}) also implies that under the worldvolume replacement (\ref{replace}) we obtain 
\be \j_2\rightarrow \j_{crit}\sin^2\theta + \j_2^{\,(eff)} = \j_{crit}\sin^2\theta +\n_{D2}\,\q_{D4}, 
\ee 
$i.e.$ \be \j_2^{\,(eff)}= \n_{D2}\,\q_{D4}.\ee
So as was the case without a black hole, 
\be   \j_1^{\,(eff)}\Leftrightarrow\j_{\zeta}, \quad \quad \j_2^{\,(eff)}\Leftrightarrow\j_{\zeta}.
\ee
Thus, in the presence of a black hole, 
(\ref{totang1}) and (\ref{totang2}) become 
\bea \j_1+ \j_1^{\,(eff)} + J&\Leftrightarrow& \j_{T}+ \j_{\zeta} +J=\J_1, \\  \j_2+ \j_2^{\,(eff)} +J &\Leftrightarrow& \j_N+\j_{\zeta}+J = \J_2. \eea
Again we have obtained worldvolume equivalents of $\J_1$ and $\J_2$.  
Moreover, the merger condition (\ref{crown}) corresponds exactly to (\ref{crown2}):
\be   (\j_1\le\j_{crit}) \Leftrightarrow (\j_T \le \d^I N_I). \ee
Overall, the level of agreement we have found is encouraging.   


\section{The Scattering Calculation:  Low Velocity Dynamical Mergers}\label{nonbps}
\subsection{Discussion of the Approximation}\label{breakdown}
Moving away from the idealized limit of an adiabatic merger, we consider the case of the supertube moving at finite but slow velocity in the BMPV background.  Here we are $not$ assuming the presence of any external torque, so, unlike the adiabatic case, $\j_2$ is conserved.  
The only relevant embedding coordinates are $\{r=r(t),\theta=\theta(t), \phi_1=\sigma,\phi_2=\phi_2(t)\}$.  The motion produces small deviations from the BPS configuration, thus breaking all supersymmetries, and the energy of the tube increases to $E_{BPS}+\Delta E$.  To calculate $\Delta E$, we treat the scattering process as motion on moduli space.  This involves expanding $\L$ to second order in the velocities $\partial_t X^i$ and the fields $\{F_{t\sigma}$, $\delta F_{tz}\},$ where $\delta F_{tz}\equiv F_{tz}-1$ is the deviation from the supersymmetric value of $\F_{tz}$.  

An important issue arises here that was not present for the adiabatic mergers.  Consider the supertube when $r^2\gg \hat{Q}$, where $\hat{Q}$ is the largest of $\{Q_{D0}, Q_{D4}, Q_{F1}\}$.  In such a situation, use of the $exact$ DBI action provides crucial insight into the behavior of a supertube of very small or very large radius $\rho_1$.\footnote{Recall that $r^2=\rho_1^2+\rho_2^2$.  When $r\rightarrow\infty$, the embedding radius $R$ satisfies $R\approx \rho_1= r\sin\theta$ and $\rho_1$ can be unambiguously termed the supertube radius.}  In the limit $\{\rho_1\rightarrow 0, \rho_2\rightarrow\infty \}$, the electric field $F_{tz}$ diverges as $1/\rho_1$; on the other hand whenever $\rho_1\rightarrow\infty$, $F_{tz}$ diminishes as $1/\rho_1$.  In either case, $\delta F_{tz}$ is no longer small compared to one.  This indicates that our moduli space approximation, including the Lagrangian (\ref{arg6}) and the quantities derived from it, can break down in certain versions of the limit $r\rightarrow\infty$.  One illustration of the breakdown is that the approximation incorrectly predicts that for motion along $\theta=\pi/2$, the effective potential $\V(r)$ approaches a constant as $r\rightarrow\infty$.\footnote{Actually, it can be shown that for a fixed value of $\theta$, $\V$ will have a term proportional to $r$ as $r\rightarrow\infty$.  Such calculations are similar to those performed for simpler backgrounds in \cite{kluson05}.}  

\begin{figure}
\begin{minipage}[t]{2.8in}
\includegraphics[width=2.8 in]{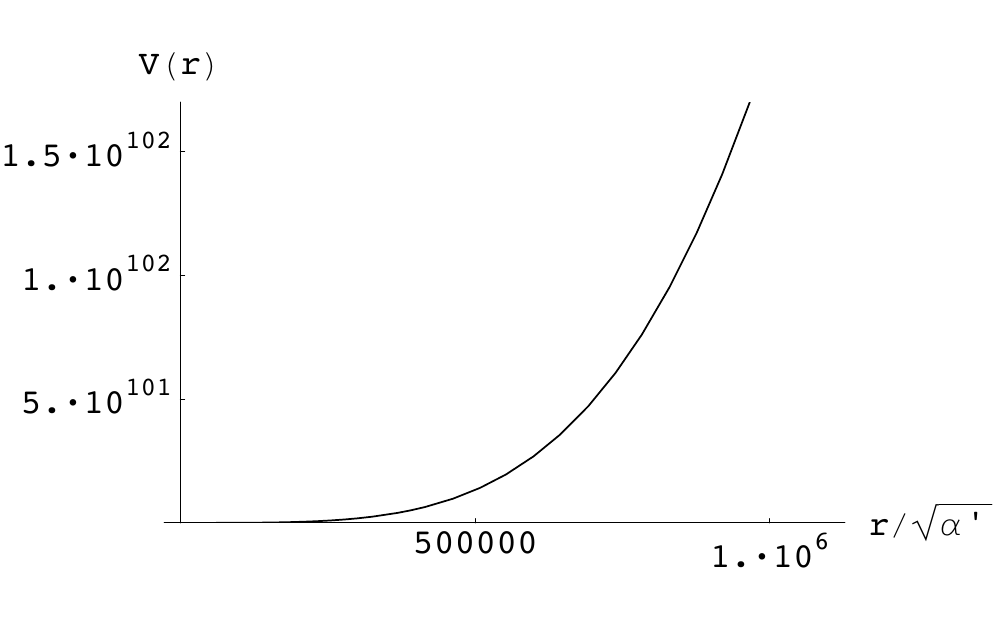} 
\caption{When $\j_1< \j_{crit}$ there is no local potential minimum.  Above, $\j_1/\j_{crit} = 0.1$.  In both figures, $\j_{crit}=2\times 10^{111}$, $\tau_{D6}V_6 = 10^{100}/\alpha'$, $Q_{D0}=10^{5}\alpha'$, $Q_{D4}=10^{6}\alpha'$, $Q_{F1}=10^{10}\alpha'$, $g_s= 10^{-5}$, $k=20$, $J=0.99\ \sqrt{N_{D0} N_{D4} N_{F1}}$, $\q_{D0}=10^{93}$, and $\q_{D4}= 10^{27}$.}
\label{nohump1}
\end{minipage}
\hfill
\begin{minipage}[t]{2.8in}
\includegraphics[width=2.8 in]{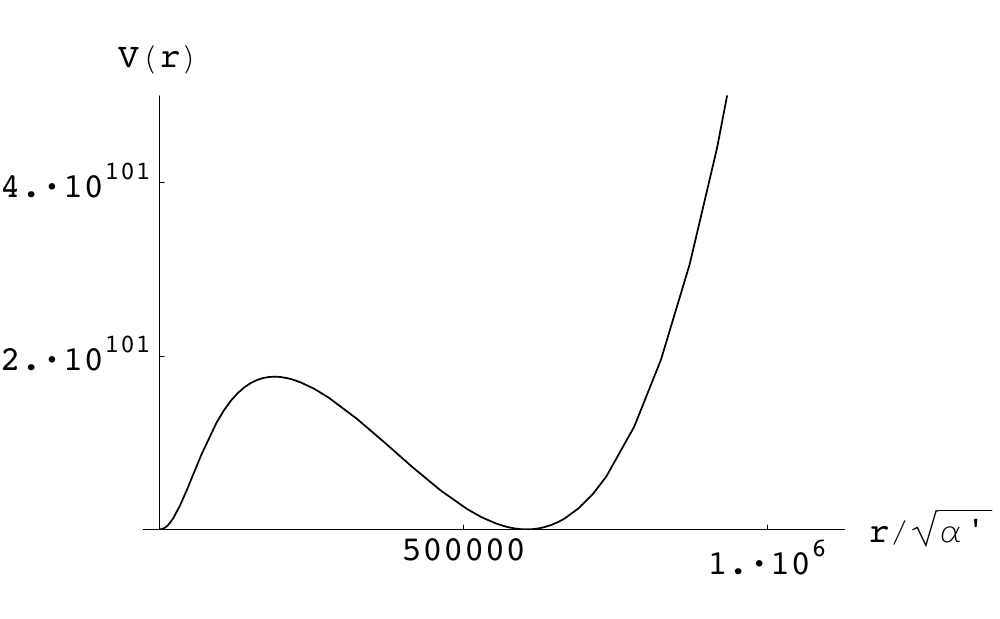}
\caption{When $\j_1> \j_{crit}$ there is a local potential minimum.  Above, $\j_1/\j_{crit} = 37.5$. In both figures $\V(r)$ is measured in units of $\frac{1}{ \sqrt{\alpha'}}$.}   
\label{hump1}
\end{minipage}
\end{figure}

However, the locations in which the approximation still holds are adequate for our purposes, and they do include a large-$r$ limit.  The moduli space Lagrangian (\ref{arg6}) can be shown to be valid in areas we will call Regions I, II and III.
Region I is the near-black hole region defined by
\be r^2\lesssim\hat{Q}. \ee    
Region II exists for the moving tubes with a BPS limit, meaning that the conditions of (\ref{jcondbps}) hold; it can overlap with Region I when $j_1<j_{crit}.$  It is given by   
\bea \label{region2} r^2 &\sim& \frac{1}{\tau_{D6} V_6 (1+B_0^2)}\frac{j_{crit}}{j_2}(j_1-j_2), \\
\{\rho_1^2 &\sim &  \frac{j_1-j_2}{\tau_{D6} V_6 (1+B_0^2)}, \,\, \rho_2^2\sim \frac{j_1-j_2}{\tau_{D6} V_6 (1+B_0^2)} \frac{j_{crit}-j_2}{j_2} \}, \nonumber \eea
in which $\rho_1$ and $\rho_2$ just take the BPS values from (\ref{rho1bps}).  Region III is specified by the limit
\be r\rightarrow\infty, \quad\quad\{\rho_1\rightarrow \hat{R},\,\, \rho_2\rightarrow\infty \},  \ee
for any radius $\hat{R}$ sufficiently close to $R_{\infty}$ of (\ref{Rinfty}).  It does not exist for the motion described in Section \ref{pi/2}, as those tubes are constrained to have $\rho_2=0$ at all times.  In the following, Figures \ref{nohump1} and \ref{vrtheta} are plots of Region I for tubes without a BPS limit, while Figures \ref{hump1} and \ref{vrtheta2} are plots of Regions I and II for tubes with a BPS limit.

There are many cases in which the domain of validity also extends between these three regions, so that a slowly moving supertube remains well described by (\ref{arg6}) as it moves from  Region I to Region III or vice versa.  All that remains is to limit ourselves to tubes for which such is the case, through the constraint $|\delta F_{tz}|  \ll1$.  With that caveat, the range described above is sufficient to capture the dynamics in the situations of interest here, which include the processes that take place near the black hole. 

\subsection{The Dynamical Lagrangian}
In anticipation of preserving the bound state of the $k$ constituent tubes, the constituents were constructed to all have the same location, embedding, charges 
and dipole charges in Section \ref{sect3}.  To continue this theme, we now require the constituents to have identical initial values of $F_{t\sigma}$, $F_{tz}$, and all components of velocity.  
Thus we ensure they will move in unison during the scattering process, and that the overall supertube maintains its integrity.

We point out that ``motion" will refer to any motion of the points on the supertube with respect to the black hole, and that this does not necessarily imply motion of the center of mass of the supertube.  This is seen through the use of the coordinates $\{\rho_1, \rho_2\}$.  Motion of the tube, when confined to the $\rho_1$-$\phi_1$ plane as in Section \ref{pi/2}, involves the case in which the center of mass of the supertube coincides with black hole, and the supertube encircles it in the $\phi_1$ direction.  The center of the tube remains stationary while the tube radius increases or decreases.  In contrast, motion in the $\rho_2$ direction \it does \rm alter the distance between the black hole and the center of mass of the tube. In fact, any motion of the supertube center of mass considered here occurs exclusively in the $\rho_2$-$\phi_2$ plane.  
As before, merging with the black hole occurs when the supertube reaches the horizon, \it i.e. \rm when 
$r=0$; additionally, $\theta > 0$ throughout the motion.

In the previous section we saw that the full supergravity solution adds nonperturbative contributions to the worldvolume values for $\q_{F1}, \j_1$ and $\j_2$.  These flux terms, $\zeta_2^c$ and $\j_{\zeta}$, are not localized to either the supertube or the black hole.  To ensure that our worldvolume analysis is applicable, we have demanded in (\ref{condbps}) that these terms are negligible compared to other supertube quantities:
\be \label{fluxterms}
\zeta_2^c = k^2\frac{\q_{D0}}{\q_{D4}}\ll \q_{F1}, \quad \quad \j_{\zeta}= k\,\q_{D0}\ll\j_{crit},\j_1\, .  
\ee
Additionally, we mention that the use of $\j_{crit}\sin^2\theta$, rather than $-\j_{crit}\cos^2\theta$, for the supersymmetric value of $\j_2$, along with the ordering $\{t,z,\sigma \}$ rather than $\{t,\sigma,z\}$, leads to results that differ somewhat  from 
those of \cite{virmani}.

With attention to these considerations, we now implement the embedding (\ref{emb}); the expanded Lagrangian density reduces from (\ref{fullarg6}) to
\bea
(\tau_{D6} k)^{-1}\L &= & - F_{z \sigma}(1+ B_0^2) +(Q_{D0} + B_0^2 Q_{D4} ) \sin^2 \theta \dot{\phi}_2+\frac {(H_{D0} +  B_0^2 H_{D4})}{2 F_{z \sigma}}F_{t\sigma}^2 \nonumber \\[2mm]
&&  + \frac{(H_{D0} +  B_0^2 H_{D4})r^2\sin^2\theta}{F_{z \sigma}}\delta F_{tz}
 + \Big( \frac {F_{z \sigma}H_{F1}}{2}  - \frac{\omega\sin^2\theta}{r^2}+
\frac{ H_{D0} H_{D4}  r^2\sin^2\theta}{2 F_{z \sigma}} \Big)\nonumber \\[2mm]
&& \times \, (H_{D0} +  B_0^2 H_{D4})  \Big(\frac {r^2\sin^2\theta}{F_{z \sigma}^2}\delta F_{tz}^2 + \dot{r}^2+r^2\dot{\theta}^2+r^2\cos^2\theta\dot{\phi}_2^2\Big).   
\label{arg6}
\eea
The $\O(v^2)$ terms in the Lagrangian are proportional to
\be \label{veloc}\Big( \frac {F_{z \sigma}H_{F1}}{2}  - \frac{\omega\sin^2\theta}{r^2}+
\frac{ H_{D0} H_{D4}  r^2\sin^2\theta}{2 F_{z \sigma}} \Big),
\ee 
and the fact that the black hole satisfies $\omega^2 < Q_{D0}\, Q_{F1}\, Q_{D4}$ is precisely the condition for this quantity to be positive for all $r$.  
Thus the metric on (the velocity subsector of) moduli space, $\tilde{g}_{ij}$, is well-behaved since we have 
\bea 
\tilde{g}_{ij}\d\dot{X^i} d\dot{X^j}& \ge & 0, \quad \mbox{where} \label{modulimet}\\
\tilde{g}_{ij}\d\dot{X^i} d\dot{X^j} &\equiv&   (H_{D0} +  B_0^2 H_{D4})\Big( \frac {F_{z \sigma}H_{F1}}{2}  - \frac{\omega\sin^2\theta}{r^2}+\frac{ H_{D0} H_{D4}  r^2\sin^2\theta}{2 F_{z \sigma}} \Big)\, v^2,
\nonumber \eea
and $v^2 = \dot{r}^2+r^2\dot{\theta}^2+r^2\cos^2\theta\dot{\phi}_2^2$ .  

The charge $\q_{F1}=k\, q_{F1}$ becomes
\be
\label{q1}
\q_{F1}= \frac{k^2 \tau_{D6}V_6 (H_{D0}+ B_0^2 H_{D4})r^2 \sin^2\theta\, }{\q_{D4}}\Bigg{(}1+ \delta F_{tz}\frac{ F_{z \sigma}^2 H_{F1}r^2 - 2\omega F_{z \sigma}\sin^2\theta+ H_{D0} H_{D4} r^4 \sin^2\theta}{F_{z \sigma}^2\, r^2}  \Bigg{)}
\ee
and the angular momenta are computed to be   
\bea
\j_1 &=& k\,j_1= \frac{\q_{F1} \q_{D4}}{k},  \\   
\label{j2eq} 
\j_2 &=& k\, j_2 = \tau_{D6} V_6 k\,   \times \Big[(Q_{D0} + B_0^2 Q_{D4} )\sin^2 \theta            \\[2mm]    
 &+& \frac{(H_{D0}+ B_0^2 H_{D4})  \dot{\phi}_2}{F_{z \sigma}} \left( F_{z \sigma}^2 H_{F1} r^2 -2\omega F_{z \sigma} \sin^2 \theta + H_{D0} H_{D4} r^4 \sin^2 \theta\right) \cos^2 \theta \Big]. \nonumber
\eea
In contrast to the supersymmetric case, $\j_2$ can now be positive or negative.  

Let us recall that $F_{ab}$ is independent of all worldvolume variables but $t$.  The field $F_{ab}$ satisfies the Bianchi identity $dF=0$, whence the relations 
\bea\label{qD4}
\partial_t F_{z \sigma} +  \partial_z F_{t\sigma}+\partial_{\sigma} F_{zt}=0 \rightarrow \partial_t F_{z \sigma}=0, \\
\partial_{t} F_{67} +  \partial_{x^7} F_{t 6}+\partial_{x^6} F_{7 t}=0 \rightarrow \partial_t F_{67}=0,\label{qD0a} \\
\partial_{t} F_{89} +  \partial_{x^9} F_{t 8}+\partial_{x^8} F_{9 t}=0\rightarrow \partial_t F_{89}=0\label{qD0b}.
\eea
Eq. (\ref{qD4}) leads to the conservation of $\q_{D4}$; the three equations (\ref{qD4}) - (\ref{qD0b}) together lead to conservation of $\q_{D0}$.    
It follows from (\ref{nd6}) and (\ref{noctcs}) that the dipole charges $\n_{D2}$ and $\n_{D6}$ remain constant as well (although for a supergravity supertube they are not conserved in general).  

Naturally, conservation of $\j_1$ and $\j_2$ follow from the equations of motion of the Lagrangian for $\phi_1$ and $\phi_2$:
\be \partial_t \frac{\partial \L}{\partial \dot{\phi_1}}=0,\quad  \partial_t \frac{\partial \L}{\partial \dot{\phi_2}}=0.\ee
(In the case of $\j_1$ we have momentarily treated $\phi_1$ as though it depended on time and then used the expanded Lagrangian (\ref{fullarg6}).)  Meanwhile, the equations of motion for $A_b$ give 
\be\label{eom1}
0=\partial_a\frac{\partial \L}{\partial F_{ab}}=\partial_t\frac{\partial \L}{\partial F_{tb}},
\ee
where $a,b$ =$\{t,\sigma,z,x^6,...,x^9\}$.  For $b=z$ this leads to the conservation of $\q_{F1}$.  When $b=\sigma$, equations (\ref{qD4}) - (\ref{qD0b}) along with (\ref{eom1}) imply that 
\be
F_{t\sigma}= const.
\ee
Thus there are trajectories in which the supersymmetric value of $F_{t\sigma}$, namely $F_{t\sigma}=0$, is maintained throughout the motion; these are the ones considered in the analysis below.  

Setting $F_{t\sigma} =0=\dot{\phi_1}$, the energy takes the form  
\bea\label{en2}
E &=& F_{tz}\frac{\partial L} {\partial F_{tz}}+\dot{r}\frac{\partial L} {\partial \dot{r}}+\dot{\theta}\frac{\partial L}{\partial \dot{\theta}}+\dot{\phi_2}\frac{\partial L}{\partial \dot{\phi_2}}-L \cr   
&=& 2\pi R_z \, \tau_{F1} \q_{F1}+\tau_{D0} \q_{D0}+V_{T^4}\,\tau_{D4} \q_{D4} + \Delta E.
\eea
After using (\ref{q1}) and (\ref{j2eq}) to eliminate $\delta F_{tz}$ and $\dot{\phi_2}$ for the conserved quantities $q_{F1}$ and $j_2$, we find that $\Delta E$, for $\theta<\frac{\pi}{2}$, is given by 
\bea  \label{deltae}
\Delta E &=& \\ 
\tau_{D6} V_6 k&\Bigg{[}&\frac {F_{z \sigma}} {2 (H_{D0} +  B_0^2 H_{D4})\,(F_{z \sigma}^2 H_{F1} r^2+ H_{D0} H_{D4}\, r^4\sin^2\theta -2\omega F_{z \sigma}\sin^2\theta)}\nonumber\\[2mm]
\times & \Bigg( &\frac{[j_1/(\tau_{D6} V_6)-(H_{D0} +  B_0^2 H_{D4})\, r^2\sin^2\theta ]^2}{\sin^2\theta} + \frac{[j_2/(\tau_{D6} V_6)-(Q_{D0}+ B_0^2 Q_{D4})\sin^2\theta]^2}{\cos^2\theta}\Bigg) \nonumber\\[2mm]
&&+  \frac {(H_{D0} +  B_0^2 H_{D4})\, (F_{z \sigma}^2 H_{F1} r^2+ H_{D0} H_{D4}\, r^4\sin^2\theta -2\omega F_{z \sigma}\sin^2\theta)}{2 F_{z \sigma} r^2} (\dot{r}^2+r^2\dot{\theta}^2) \Bigg{]}. \nonumber
\\ \nonumber
\eea
The form of $\Delta E$ when $\theta=\frac{\pi}{2}$ is presented below.  Again, the supertube always satisfies $\theta > 0$.

\subsection{Scattering in the Plane $\theta=\frac{\pi}{2}$}   \label{pi/2}
The simplest motion of the supertube is that confined to constant values of $\theta$ and $\phi_2$.  However, the only trajectories of constant $\theta$ allowed by the equations of motion 
are for $\theta=\pi/2$ \cite{virmani}.  
In this plane, $\dot{\theta}$ vanishes and the $\phi_2$ coordinate of the supertube is undefined.  The motion is purely in the $\rho_1$-$\phi_1$ plane, and since $\rho_2=0$, $r$ itself reduces to $\rho_1$.  Such motion is merely the supertube changing its radius while its center of mass, coinciding with the black hole, remains motionless.  We will not consider the behavior of the effective potential $\V(r)$ at very large $r$ since the moduli space approximation breaks down for large $\rho_1$, as noted above.  

The angular momentum $\j_2$ has a conserved nonzero value of $\j_2 = \j_{crit}$. 
The excess energy $\Delta E$ simplifies to 
\bea \label{deltaer}
&&\Delta E|_{\theta=\frac{\pi}{2}} \\
&& = \tau_{D6} V_6\, k\, \frac {(H_{D0} +  B_0^2 H_{D4})\, (F_{z \sigma}^2 H_{F1} r^2+ H_{D0} H_{D4}\, r^4 -2\omega F_{z \sigma})}{2 F_{z \sigma} r^2} \dot{r}^2 +\V(r),\nonumber
\eea
where $\V(r)$ is given by
\bea \label{eff1}
\V(r)&=&\Delta E|_{\dot r=0,\mbox{ }\dot{\theta}=0,\mbox{ }\theta=\frac{\pi}{2}}  \nonumber\\[2mm]
&=&  \tau_{D6} V_6\, k\, \frac {F_{z \sigma}\, r^2\,[j_1/(\tau_{D6} V_6) - (H_{D0} +  B_0^2 H_{D4})\, r^2 ]^2} {2 r^2 (H_{D0} +  B_0^2 H_{D4})\,(F_{z \sigma}^2 H_{F1} r^2+ H_{D0} H_{D4}\, r^4 -2\omega F_{z \sigma})}\,.
\eea
The denominator of $\V(r)$ is assured of being positive due to the inequality (\ref{modulimet}); meanwhile, keeping $r^2$ explicit in the numerator makes clear the fact that $\V(r)$ vanishes at the origin.   

The function $\V(r)$ vanishes as $r\rightarrow 0$.  However, if the expression in brackets in (\ref{eff1}) vanishes, the potential also goes to zero at another value of $r$, thus creating a local minimum.  The latter occurs when
\bea
(H_{D0} +  B_0^2 H_{D4}) r^2 &= & Q_{D0}+r^2+B_0^2(Q_{D4}+r^2)=\frac{j_1}{\tau_{D6} V_6} \label{bps_r.2} \\
\rightarrow \tau_{D6} V_6\,(1+B_0^2)r^2 
 &=& j_1 - j_{crit}.  
\eea
Due to the form of (\ref{eff1}), $\V=0$ at this stable minimum, and it occurs  when
\be
j_1 >  j_{crit},	\label{nocrown}
\ee
at some $r_1>0$ where
\be	\label{r_1}
r_1^2 =\frac{j_1 - j_{crit}}{\tau_{D6} V_6  (1+B_0^2)} \, .
\ee  
When $j_1\le j_{crit}$ the potential is attractive for all $r$, with no impediment to merging.  

As noted in Section \ref{embed}, in an adiabatic merger, supertubes with $j_1>j_{crit}$ cannot merge with the black hole.  We see here that dynamical mergers differ crucially in that such tubes only encounter a finite potential barrier, and thus a merger is possible even when $j_1>j_{crit}$.  
Further inspection reveals that (\ref{bps_r.2}) is just (\ref{bps_r}), and (\ref{r_1}) is (\ref{r}), which gives the location of a BPS supertube, for $\theta=\pi/2$.  This is consistent with the fact that a motionless supertube with $\V=0$ saturates the BPS bound and is thus in a BPS configuration.

\subsection{Scattering for $\theta<\frac{\pi}{2}$} 
When $\theta<\pi/2$, we use (\ref{deltae}) to arrive at the effective potential
\bea \label{eff2}
\V(r,\theta)&=&\Delta E|_{\dot r=0,\dot\theta=0} \cr
&=&  \tau_{D6} V_6 \,k\,\frac {F_{z \sigma}r^2} {2 r^2(H_{D0} +  B_0^2 H_{D4})\,(F_{z \sigma}^2 H_{F1} r^2+ H_{D0} H_{D4}\, r^4\sin^2\theta -2\omega F_{z \sigma}\sin^2\theta)}  \\[2mm]
&&\times \Bigg{(} \frac{[j_1/(\tau_{D6} V_6)- (H_{D0} +  B_0^2 H_{D4})\, r^2\sin^2\theta ]^2}{\sin^2\theta} + \frac{[j_2/(\tau_{D6} V_6) - (Q_{D0}+ B_0^2 Q_{D4})\sin^2\theta]^2}{\cos^2\theta}\Bigg{)}. \nonumber
\eea
\begin{figure}
\begin{minipage}[t]{2.8 in}
\includegraphics[width=3.1 in]{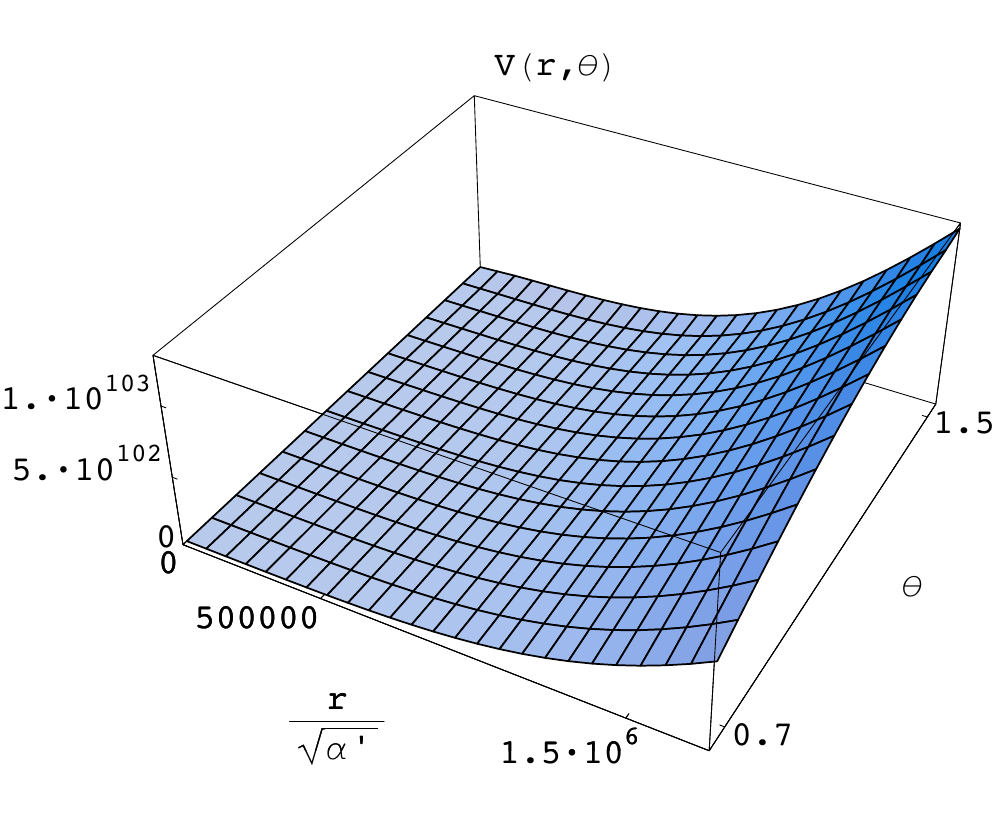}
\caption{$\V(r,\theta)$ with no potential barrier.  Above, $\j_1/\j_{crit} = 0.1$ and $\j_2/\j_{crit} = 0.05$.  In both figures, $\j_{crit}= 2\times10^{111}$, $\tau_{D6}V_6 = 10^{100}/\alpha'$, $Q_{D0}=10^{5}\alpha'$, $Q_{D4}=10^{6}\alpha'$, $Q_{F1}=10^{10}\alpha'$, $g_s= 10^{-5}$, $k=20$, $J=9.9\times 10^{125}$, $\q_{D0}=10^{93}$, $\q_{D4}= 10^{27}$ and $\V$ is in units of $\frac{1}{ \sqrt{\alpha'}}$. }\label{vrtheta}
\end{minipage}
\hfill
\begin{minipage}[t]{2.8 in}
\includegraphics[width=3.1 in]{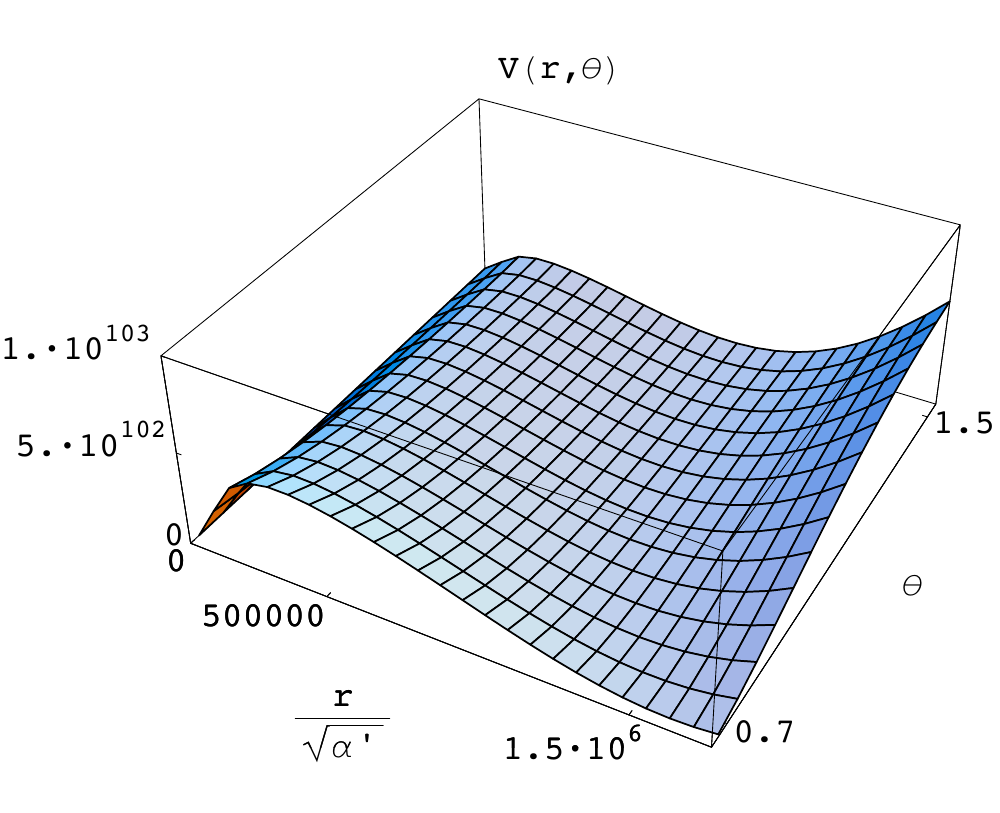}
\caption{$\V(r,\theta)$ with the same parameters as in the previous figure, except that $\j_1/\j_{crit} = 100$ and $\j_2/\j_{crit} = 0.5$.  There is now a potential barrier for all $\theta$ (only a portion of which can be shown above).} 
\label{vrtheta2}
\end{minipage}
\end{figure}
Like $\V(r)$ in (\ref{eff1}), $\V(r,\theta)$ is nonnegative.  Now, our analysis has assumed that 
\bea \label{del} & |\delta F_{tz}| & \ll1,
\\  v_{\phi_2}\equiv & |r\cos\theta\dot{\phi_2}| & \ll 1.\label{vel}\eea 
The constraint (\ref{del}) implies that for a given supertube there is a lower range of $\theta$ below which our approximations break down; meanwhile, (\ref{vel}) invalidates our approximations for upper range of $\theta$ (and sometimes a lower range also).  
There are other supertubes, naively acceptable, whose combinations of $\j_1$  and $\q_{D4}$ or $\j_2$ and  $\q_{D4}$ are such that (\ref{del}) or (\ref{vel}), respectively, excludes them from our analysis for any value of $\theta$.  

Similarly to $\V(r)$ in Section \ref{pi/2}, $\V(r,\theta)$ vanishes at $r=0$, independently of $\theta$.  Does it contain a local minimum as well?  To explore this possibility we examine the vanishing of $\V$.  The first term in brackets in (\ref{eff2}) always vanishes on some curve in the $r$-$\theta$ plane.  
The second bracketed term can vanish only if $0< j_2\le j_{crit}$; if such is the case it does so at the value $\theta=\theta_1$ where
\be
\sin\theta_1 = \sqrt{\frac{j_2}{j_{crit}}}\,\,.
\ee
Clearly, for $\V(r,\theta)$ to vanish when $r\neq 0$, both bracketed quantities of (\ref{eff2}) must vanish at the same location.  
This is possible when 
\be\label{j2criterion}  
0< j_2\le j_{crit}, \quad j_1>j_2.
\ee
It follows that when (\ref{j2criterion}) holds, $V(r,\theta)$ can indeed vanish, and this location constitutes a stable minimum.  This minimum occurs at $(r_1,\theta_1)$ with $r_1$ given by 
\be\label{r1full}
r_1^2=\frac{1}{\tau_{D6} V_6 (1+B_0^2)}\frac{j_{crit}}{j_2}(j_1-j_2).
\ee
Equation (\ref{j2eq}) tells us that when $\theta=\theta_1$ the supertube has no velocity in the $\phi_2$ direction, $i.e.$ $\dot{\phi_2}=0$.

Since a motionless supertube with $\V=0$ is in a BPS configuration, we expect that the conditions for $\V=0$ to arise mirror the relations found in earlier sections for the BPS supertube.  Such is precisely the case:  (\ref{j2criterion}) is just (\ref{jcondbps}), (\ref{r1full}) is identical to (\ref{rbps}), and the vanishing of the first and second bracketed terms of  $\V(r,\theta)$ corresponds to the conditions (\ref{bps_r}) and (\ref{j2real}) respectively.  It follows, of course, that $\rho_1$ and $\rho_2$ take the BPS values of (\ref{rho1bps}) and thus that this local minimum occurs at the location that specified Region II in (\ref{region2}).  Since we would not expect any non-BPS configurations to be in equilibrium with the black hole, it makes sense that we find no other true local minima in this potential.
 
The presence of the potential barrier when $\theta<\pi/2$ is a $\theta$-dependent phenomenon.  Specifically, the vanishing of $\partial_r \V$ at two nonzero locations does not require the simultaneous vanishing of $\partial_{\theta} \V$.  A potential barrier for motion in the $r$ direction in the cross sections $\V(r,\theta=const.)$ appears for certain ranges of $\theta$ even when (\ref{j2criterion}) is $not$ satisfied.  Typically, the barrier is present for all $\theta$ except the values $\theta_a<\theta<\theta_b$, where $\theta_a$ and $\theta_b$ are complicated functions of the parameters.  Figures \ref{vrtheta} and \ref{vrtheta2} provide examples of cross sections with and without a potential barrier.  The barrier is present for $all$ $\theta$ when $j_1/j_{crit}$ is sufficiently large, as it is for Figure \ref{vrtheta2}; unexpectedly, this too can occur when (\ref{j2criterion}) does not hold.  The criteria for such a barrier to occur are complicated and $\theta$-dependent.  What is guaranteed by the conditions of (\ref{j2criterion}) is that there will at least be a barrier for some values of $\theta$.
 
As concluded in \cite{virmani}, supertubes in BPS configurations 
are in positions of stable equilibrium, separated from the black hole by a potential barrier.  It is possible to overcome the barrier with sufficient kinetic energy, or with the application of additional forces to the supertube (it can also quantum mechanically tunnel through the barrier).  We also emphasize that the actual dynamical behavior of this section is independent of $k$, the number of constituent D6-branes; the supertube behaves the same as would any of its constituents individually.  For completeness we mention that the analogous condition to (\ref{j2criterion}) for the D0-F1 supertube is
\be
0< j_2\le N_{D4}, \quad  j_1>j_2.
\ee
 
\section{Results of the Dynamical Merger}\label{overspin}  
\subsection{The Subcritical Black Hole} \label{sec61}
The BMPV black hole satisfies $J_1=J_2=J$ where $J_i$ is its angular momentum in the plane of $\phi_i$; moreover, $J$ is constrained by \be\label{bound}
J^2< N_{D0} N_{D4} N_{F1}. 
\ee
Violation of an angular momentum bound we will refer to as ``overspinning" a black hole.\footnote{``Overspin" here merely denotes exceeding the angular momentum bound, resulting in some (as yet undetermined) black object; we are not suggesting that naked CTCs would actually be created because the second law of thermodynamics prevents the occurrence of such causally sick geometries.}  
The ADM mass and entropy of the BMPV black hole are given by 
\begin{eqnarray}
M_{BMPV} &=& \frac{\pi}{4 G_5}\left( Q_{D0} + Q_{D4} + Q_{F1} \right) = \frac{1}{g_s\sqrt{\alpha'}}
\left(N_{D0} + \frac{\l^4}{\alpha'^2}N_{D4} + \frac{g_s R_z}{\sqrt{\alpha'}}N_{F1} \right), 
\label{bmpvmass}\\[2mm]
S_{BMPV} &=& \frac{2\pi^2}{4 G_5} \sqrt{Q_{D0} Q_{D4} Q_{F1}  -
\omega^2} = 2 \pi \sqrt{N_{D0} N_{D4} N_{F1} - J^2}. \label{bmpvent}
\end{eqnarray}
The merger process we consider keeps $J_1 = J_2$ by letting two identical supertubes simultaneously merge with 
the black hole, where one tube has its circumference in the $\phi_1$ direction and the other in the $\phi_2$ direction.  We limit ourselves to the case in which 
the $\phi_1$-wrapped tube moves in the $\theta=\pi/2$ plane and the $\phi_2$-wrapped tube moves in the $\theta=0$ plane, so that the analysis of Section \ref{pi/2} applies to both tubes.  During this process, the centers of the tubes remain motionless while their radii shrink until the merger occurs.  We let both tubes carry angular momentum $\j-\j_{crit}$ along their circumferences, where $\j_{crit}\ll j$, while carrying angular momentum $\j_{crit}$ in their transverse planes.  Thus the black hole receives an angular momentum contribution of $+\j$ in both the $\phi_1$ and $\phi_2$ planes and we write its change as 
\be  \j =\frac{\q_{F1}\, \q_{D4}}{k}\,+\j_{crit}.
\ee 
The requirements (\ref{fluxterms}) on the flux terms ensure that we can safely neglect their contribution to the post-merger black hole.  Thus the changes in the black hole parameters are just $\Delta J=\j$, $\Delta N_{I}=2\q_{I}$.  For ease of computation, we will focus on the case in which $\j_{crit}\ll \j$ so that $\j\approx\q_{F1}\, \q_{D4}/k$ to an excellent approximation.  Hence, we are assuming that
\be
k\,\q_{D0} \ll \j_{crit}\ll \j \ll J, \quad k^2\frac{\q_{D0}}{\q_{D4}}\ll \q_{F1},\quad \q_{I} \ll N_{I}, 
\quad |\delta F_{tz}|\ll 1,   \label{cond2}	\ee
so that we remain within the range of validity of our approximations.  We recall that motion at constant $\theta$ cannot be adiabatic, and observe that the restriction $\j_{crit}\ll \j$ compels these mergers to have a potential barrier. 

Now, in a dynamical merger the presence of the energy $\Delta E$, even if it is very small, implies that the final state of the merger is not BPS and thus $not$ a BMPV black hole.  Therefore the entropy of the end product will differ from (\ref{bmpvent}).   
One candidate for the result of the merger is the non-BPS, charged, rotating black hole found by Cvetic and Youm (CY) in \cite{cy1,cy2}.  This solution in general has $J_1\neq J_2$.  The post-merger black hole parameters we will write as $J', N'_I, Q'_I$, $etc$.  The CY black hole in the near-BPS limit has an ADM mass and difference in angular momenta given by the following:
\bea \label{cymass} M'_{CY}  &=&  \frac{\pi}{4 G_5} \left(Q'_{D0}+Q'_{D4}+Q'_{F1}+\frac{m^2}{2} \left(\frac{1}{Q'_{D0}} + \frac{1}{Q'_{D4}} + \frac{1}{Q'_{F1}}\right) + \O(m^4)\right), \\
| J'_1 - J'_2 | &\le& \frac{\pi m}{ 2 G_5} \sqrt{Q'_{D0}\, Q'_{D4}\, Q'_{F1}}  \left(\frac{1}{Q'_{D0}} + \frac{1}{Q'_{D4}} + \frac{1}{Q'_{F1}}\right) + \O(m^2), \label{jdiff} 
\eea
(correcting a typographical error in Eq. (2.12) of \cite{virmani}), where $m$ parameterizes energy above extremality and satisfies $m\ll Q'_{D0}, Q'_{D4}, Q'_{F1}$.  When $J'_1=J'_2=J'$ the CY entropy has the form
\be \label{cyent}  S_{CY} =\frac{2\pi^2}{4G_5}\left(\sqrt{Q'_{D0} Q'_{D4} Q'_{F1}  -
\omega'^2} + m\frac{Q'_{D0}Q'_{D4}+Q'_{D0}Q'_{F1}+Q'_{D4}Q'_{F1}}{\sqrt{Q'_{D0}Q'_{D4}Q'_{F1}}} + \O(m^2)\right), 
\ee  
where $(\pi/4 G_5)\,\omega'=J'$.  
Since after the merger the energy above the BPS bound is just $\Delta E$, we can use (\ref{cymass}) to relate $\Delta E$ to $m$:
\be \label{deltaem}\Delta E=  \frac{\pi}{4 G_5} \frac{m^2}{2} \left(\frac{1}{Q'_{D0}} + \frac{1}{Q'_{D4}} + \frac{1}{Q'_{F1}}\right)  + \O(m^4). 
\ee

It is also convenient to give the expression for the change in entropy between a CY black hole with angular momenta equal to $J'$ and charges $\{Q'_{D0}, Q'_{D4}, Q'_{F1}\}$, and a hypothetical BMPV black hole with the same parameters:
\be \label{diffent} \delta S^{(1)} =S'_{CY}-S'_{BMPV}= \frac{2\pi^2}{4G_5}\left(m\frac{Q'_{D0}Q'_{D4}+Q'_{D0}Q'_{F1}+Q'_{D4}Q'_{F1}}{\sqrt{Q'_{D0}Q'_{D4}Q'_{F1}}} + \O(m^2) \right). \ee
In terms of $\Delta E$, the constraint on angular momentum for a post-merger CY black hole with $J'_1= J'_2$ is \cite{virmani}   \begin{eqnarray} \label{cybound}
J^{'\,2} &<& N'_{D0} N'_{D4} N'_{F1}    \\ &\times& \Bigl( 1 + \Delta E
\frac{2 G_5 }{\pi} \frac{Q_{D4}^{'\,2} Q_{F1}^{'\,2}+ Q_{D0}^{'\,2} Q_{F1}^{'\,2}
+Q_{D4}^{'\,2} Q_{D0}^{'\,2} + 2 Q'_{D0} Q'_{D4} Q'_{F1}(Q'_{F1} +
Q'_{D0} + Q'_{D4})}{ Q'_{D0} Q'_{D4} Q'_{F1} (Q'_{D4} Q'_{F1}
+ Q'_{D0} Q'_{F1} + Q'_{D4} Q'_{D0})} \nonumber \cr &+&
\O(\Delta E)^2 \Bigr).
\end{eqnarray}
Reference \cite{virmani} discussed the process of using external forces to gently push the pair of supertubes past the potential barrier, and then moderate their collapse into the black hole so that the end product of the merger would be just above extremality.  It follows that $m$ and therefore $\Delta E$ can be (for classical purposes) arbitrarily small. 

However, the assumption that the end-product of the merger will merely be a CY black hole may not always be appropriate.  Strikingly, it can be seen using (\ref{bmpvent}) that adding certain supertubes to the black hole appears to involve a $decrease$ in the BMPV entropy.  Such a situation occurs when
\bea  
S'_{BMPV } & < & S_{BMPV}, \quad i.e. \nonumber \\[2mm]
 N'_{D0} N'_{D4} N'_{F1} - J^{'\,2}  &<& N_{D0} N_{D4} N_{F1} - J^2, \label{decrease0}
\eea
which, for $\j\approx\q_{F1}\, \q_{D4}/k$, happens when the charges of the supertubes and original black hole satisfy\footnote{It is not obvious, but such mergers are possible while all the conditions of (\ref{cond2}) still hold.}
\be \j\approx\q_{F1}\, \q_{D4}/k >
\frac{N_{D0}N_{D4}\q_{F1}+N_{D0}N_{F1}\q_{D4}+N_{D4}N_{F1}\q_{D0}}{J}\, . 
\label{decrease1}   \ee  
This would appear to contribute a negative change in the entropy:
\bea \label{decrease2} \Delta S^{(2)} &=& S'_{BMPV } - S_{BMPV} \\
&\approx& -2\pi \frac{J \q_{F1}\q_{D4} - k( N_{D0}N_{D4}\q_{F1}+N_{D0}N_{F1}\q_{D4}+N_{D4}N_{F1}\q_{D0})}{k \sqrt{N_{D0} N_{D4} N_{F1} - J^2}}.   \nonumber
\eea    
It is emphasized that $\Delta S^{(2)} $ is a change representing a dynamical process, while $\delta S^{(1)}$, as written, is simply the difference in the expressions for the CY and BMPV entropies.  

Now, the entropy of the supertube itself is completely negligible relative to that of the black hole \cite{warner2}.\footnote{And since the black hole has no dipole charges, there is no danger of the `entropy enhancement' discussed in \cite{bena5}.}  Let us perform the merger (involving supertubes that satisfy (\ref{decrease1})) in such a way that the same would be true of any radiation produced.\footnote{The role of radiation in this scenario is discussed in detail in \cite{virmani}.}  If the outcome were a CY black hole, the overall change in entropy would essentially be 
\be \Delta S = S'_{CY} - S_{BMPV} = \delta S^{(1)}+\Delta S^{(2)}. \ee 
But from (\ref{deltaem}) and (\ref{diffent}), we see that if $\Delta E$ were small enough, the entropy difference $\delta S^{(1)}$ would not be enough to offset the decrease $\Delta S^{(2)}$, $i.e.$ the overall change in entropy $\Delta S$ would appear to be negative, in violation of the generalized second law (GSL) of thermodynamics.  Therefore, the GSL prevents the outcome of this merger from being merely a CY black hole.  If, as conjectured, the CY solution is the only solution of minimal five dimensional supergravity that reduces to the BMPV black hole in the BPS limit, it seems that $no$ black hole is a suitable candidate for the end product of such a merger.\footnote{On the other hand, in \cite{reall3} it was suggested that other non-extreme black hole  solutions with less symmetry than CY exist; if such solutions were stable these would also have to be considered.}  
Instead, other black objects and combinations of them must be considered, and in \cite{virmani} it was suggested that the black hole could fragment into a pair of black rings, or a black ring encircling a rotating black hole.  Some properties of black rings are given in Appendix \ref{appc}.  

The above analysis of the entropy extends the previous notion of black hole fragmentation.  That is, the claim of \cite{virmani} was made only in the context of overspinning a near-critical black hole; here we see that it must be considered for certain subcritical black holes also.  Thus we have presented a larger class of \it fragmentation mergers. \rm  Reference \cite{bena2} discussed adiabatic supertube/BMPV  mergers as a special case of adiabatic black ring/BMPV mergers.  Their results raise the interesting possibility that, if fragmentation mergers indeed occur, the post-radiation endpoint for the above fragmentation process could be a pair of concentric perpendicular black rings surrounding a rotating black hole, $i.e.$ a charged version of the ``bi-ring black Saturn" of  \cite{elvang3}.  The investigations in \cite{gauntlett1} have shown that for overspinning mergers, discussed in more detail below, a pair of black rings without a central black hole is also an option.  An in-depth exploration of the nature of this final state would certainly be illuminating.

\subsection{The Near-Critical Black Hole: $J^2\approx N_{D0} N_{D4} N_{F1}$}\label{sec62}
When the BMPV angular momentum is very near 
the critical value $\sqrt{N_{D0} N_{D4} N_{F1}}$, we must consider the question of whether or not it is possible for supertube mergers to overspin the black hole.  Since fragmentation mergers satisfy 
\be N'_{D0} N'_{D4} N'_{F1} - J^{'\,2}  < N_{D0} N_{D4} N_{F1} - J^2, \nonumber \ee
while overspinning mergers by definition have \be \label{oversp} N'_{D0} N'_{D4} N'_{F1} - J^{'\,2}  < 0, \ee 
it is clear from (\ref{bound}) that overspinning mergers are just a subset of fragmentation mergers.

\begin{figure}
\includegraphics[width=3.5in]{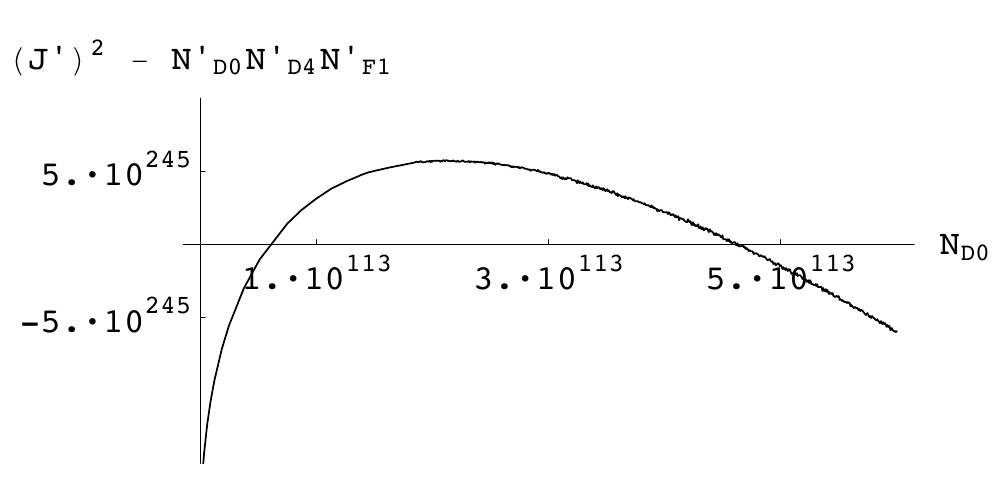}
\caption{This figure shows the range of $N_{D0}$ for which the BMPV bound is violated when $N_{F1}=10^{100}, N_{D4}=10^{46}, \q_{F1}=10^{86}, \q_{D4}=5\times 10^{32}, \q_{D0}=10^{100}$, $k=90$  and $J\approx \sqrt{N_{D0} N_{D4} N_{F1}}$.  Here $N^{spin}_{-} = 5.98\times 10^{112}$ and 
$N^{spin}_{+} = 4.64\times 10^{113}. $}
\label{ov1}  
\end{figure}

Let us continue our analysis of the process that involves a pair of identical supertubes merging with the black hole, as described in Section \ref{sec61}.  Straightforward algebra reveals that under certain conditions it is indeed possible for such a merger to produce a black object which satisfies (\ref{oversp}).  To delineate these conditions, we define
\bea
&N^{spin}_{\pm}&\equiv \frac{N_{F1} N_{D4}}{2(\q_{F1} N_{D4}+\q_{D4} N_{F1})^2} \\
&\times & \Big(\j^{\, 2} -2 \q_{D0}\, ( \q_{F1} N_{D4} + \q_{D4} N_{F1}) \pm \j\sqrt{\j^{\, 2} -4\q_{D0}( N_{D4}\,  \q_{F1} + N_{F1}\, \q_{D4})}\, \,\Big).\nonumber
\eea
If $N^{spin}_{-}\gg q_{D0},$ 
then a criterion to violate the BMPV bound (\ref{bound}) takes the form
\be \label{overspin1a}
N^{spin}_{-} <  N_{D0} < N^{spin}_{+}\, ;
\ee
otherwise (\ref{overspin1a}) merely becomes 
\be \label{overspin1b} \q_{D0}\ll N_{D0} < N^{spin}_{+}\, . \ee 
An example is shown in Figure \ref{ov1}.  We note that $N^{spin}_{\pm}$ depends on parameters of both the supertube and the black hole.  

Now, for $N^{spin}_{\pm}$ to be real, the quantity in the square root must of course be nonnegative; it turns out that violation of the bound requires that it be positive.  So in addition it is required that
\be\label{overspin2}
\j^{\, 2}-4\q_{D0}\, (N_{D4}\,  \q_{F1} + N_{F1}\, \q_{D4})>0
\ee
for the BMPV bound to be exceeded.  The excess energy $\Delta E$ can be arbitrarily small, and it is clear that if we perform the merger such that $\Delta E$ is less than a certain value, the CY angular momentum bound (\ref{cybound}) will be violated also.  Hence, a pair of identical supertubes that satisfy (\ref{overspin1a}) [or (\ref{overspin1b}) as appropriate] and (\ref{overspin2}) can indeed 
overspin a near-critical BMPV black hole.  

For overspinning mergers involving a pair of D0-F1 supertubes (subject to the relevant restrictions analogous to (\ref{cond2})), instead of the set (\ref{overspin1a} - \ref{overspin2}) the only requirement for an overspin is
\be N_{D4}< \frac{N_{D0}N_{F1} \, j^2}{(N_{F1} q_{D0} + N_{D0} q_{F1})^2}; \ee 
there is no analog of $N^{spin}_{-}$.  There is also a D0-F1 counterpart to (\ref{decrease1}):
\be j\approx q_{F1}q_{D0} > \frac{N_{D4}(N_{D0}q_{F1}+N_{F1}q_{D0})}{J}  \,.
\label{decrease3}   \ee
It is straightforward to show that overspinning mergers must produce a potential barrier for these tubes as well.

Thus we have shown that, in the moduli space approximation, it $is$ possible to violate the BMPV bound, but such mergers, like all fragmentation mergers discussed here, must produce a potential barrier (just as was found in \cite{virmani} for a pair of D0-F1 supertubes).  We have assumed that the process involves only perturbations small enough to make no perceptible change in the relations (\ref{overspin1a} - \ref{overspin2}).  The excess energy $\Delta E$ can be arbitrarily small, and it is clear that if we perform the merger such that $\Delta E$ is less than a certain value, the CY angular momentum bound (\ref{cybound}) will be violated also. 

Another process of interest is to let a single supertube, or two supertubes with sufficiently unequal angular momenta, merge with the black hole.  Such a merger can be arranged to have $m$ low enough so that the resulting object cannot satisfy the constraint (\ref{jdiff}).  (Recall that in the adiabatic merger, this would not happen because the applied torque increases $\j_2$ to ensure that $\j_1= \j_2$ at the point of the merger.)  Violation of (\ref{jdiff}) shows that the end result of these mergers is neither a BMPV nor CY black hole.  Such phenomena, like overspinning mergers, are special cases of fragmentation mergers. 


\section{Conclusions}\label{discussion}
The usefulness of the DBI description is that it allows us to treat time-dependent phenomena, and the success of this approach for the two-charge supertube motivated the present study.  It is perhaps unsurprising that the basic attributes of the scattering process for three-charge supertube -- the existence of a critical angular momentum, the existence of a stable minimum under certain conditions, the necessity of a potential barrier for any violation of the BMPV bound to take place -- are very similar to those for the the two-charge supertube.   

Meanwhile, we have established a correspondence with the supergravity results of \cite{bena2}, thereby  identifying and addressing a gauge ambiguity in the background R-R fields.  Such a comparison could, of course, be pursued further, in hopes of understanding the origin of the physics underlying the replacement (\ref{replace}).  It should be kept in mind that at the time of writing, the various worldvolume formulations of the three-charge supertube from \cite{elvang} and \cite{bena} are still distinct.  Reference \cite{elvang} presented a worldvolume description of a  calibrated supertube, and it would be interesting to see if such a tube has a different merger outcome.  And of course, a fair amount could be clarified about D0-D4-F1 (and thus D1-D5-P) microstates, in the process of showing explicitly in any of these formulations that the three-charge supertube indeed has a discrete spectrum, in the manner of \cite{marolf2, ohta2}.

The notion of a thermodynamic instability of (small deformations of) a BMPV black hole was brought up in \cite{virmani};\footnote{We say `deformation' because strictly speaking, the instability was posited for CY, not BMPV, black holes.  This is because the latter, being BPS, are not expected to have linear instabilities.  Again, it is assumed here that there are no small deformations of the BMPV black hole other than the CY solution.} it was suggested that an overspinning merger could cause the black hole to fission into several black objects.  Here we have extended this proposal.  Due to the generality of (\ref{decrease3}) for the two-charge case and (\ref{decrease1}) for the three-charge case, overspinning mergers are but a \it subset \rm of fragmentation mergers, because (\ref{decrease1}) and (\ref{decrease3}) hold for many supertubes that do not overspin the black hole.  Therefore, the supposed instability arises under more general conditions than discussed in \cite{virmani}.  

Candidates for the endpoint of a fragmentation merger of the type described in Section \ref{overspin} would presumably include slightly nonextremal versions of the configurations found in \cite{gauntlett2}, $i.e.$ a pair of orthogonal charged black rings in the overspinning case, or a pair of charged black rings with an additional central CY black hole in the non-overspinning case.  These might be realized by charged versions of black bi-rings found in \cite{elvang3, izumi} and the bi-ring black Saturn proposed in \cite{elvang3}.  Recent attempts to charge black Saturn, however, have been plagued with singularities \cite{mann}.  Moreover, indications from an analysis of the exact DBI action are that fragmentation mergers might not occur after all.  This topic is currently under investigation.  A detailed study of this process will have implications for our knowledge of the phase structure of higher dimensional black objects \cite{figueras3, emparan2} in the charged regime. 


\subsection*{Acknowledgments}  
I extend my gratitude to Don Marolf, Amitabh Virmani, and Per Kraus for many valuable clarifying discussions.  I have also benefited from the input of Iosif Bena, Joe Polchinski, and Tristan Hubsch.  Helpful suggestions and encouragement were provided by Lisa Dyson and Aaron Roane.  The support of the UCSB Graduate Division through the GRIP Program is much appreciated, as is assistance from the Howard University Department of Physics and Astronomy, and the Howard University Graduate School.  Further support has come from US Department of Energy grant DE-FG02-94ER-40854.  Finally, I would like to acknowledge the hospitality of the Kavli Institute for Theoretical Physics at UCSB, where part of this work was completed. 

\appendix
\section{The General Second-Order Lagrangian}  \label{appa}
The action $\S$ satisfies
\be
 \S = \int \L\, d^7 x  =\int {(\L_{BI}+\L_{WZ})}\, d^7 x, 
\ee 
where $\L$ is the Lagrangian density.  Our use of the moduli space approximation involves the expansion of $\L$ to second order in the velocities $\partial_t X^i$ and the gauge fields $\{F_{t\sigma}, \delta F_{tz}\},$ where $\delta F_{tz}= F_{tz}-1$, using the fields given in Section \ref{background} and following the methods outlined in the Appendix of \cite{virmani}.  It must be kept in mind that this approximation can break down when $r^2 \gg \hat{Q}$ unless the supertube motion as $r\rightarrow\infty$ is from Region I to Region III, as discussed in Section \ref{breakdown}.  The result is that
\bea \label{fullarg6}
&(&\tau_{D6} \,k)^{-1}\L= -F_{z \sigma}(1+  B_0^2)-\Big(Q_{D0}+B_0^2\,Q_{D4}\Big)\, \psi_{t\sigma} \sin^2\theta\cr
&&+\frac {F_{z \sigma}}{2}\Big(H_{D0} H_{F1} \Delta_{tt} + B_0^2 \,(H_{F1} H_{D4} \Delta_{tt})\Big)+ 
\frac {F_{z \sigma}^{-3}}{2} H_{D0} H_{D4}\,(H_{D0} + B_0^2 H_{D4})\,\Delta_{\sigma\sigma}^2\, \delta F_{tz}^2 \cr
&& +F_{z \sigma}^{-2} (H_{D0} + B_0^2 H_{D4})\,( H_{D0} H_{D4} \Delta_{\sigma t} - \gamma_{\sigma}\,\delta F_{tz})\,\Delta_{\sigma\sigma}\, \delta F_{tz}  \cr
&& +(H_{D0} + B_0^2 H_{D4})\,(\Delta_{\sigma t}-\gamma_{\sigma}\Delta_{tt}+H_{F1}\Delta_{\sigma t}\delta F_{tz} )\cr
&&+\frac {F_{z \sigma}^{-1}}{2}(H_{D0} + B_0^2 H_{D4})\, \Big(F_{t\sigma}^2+ H_{D0} H_{D4} \Delta_{\sigma\sigma}\Delta_{tt}
 +\delta F_{tz}\, (2\Delta_{\sigma\sigma}-4\gamma_{\sigma}\Delta_{\sigma t} \cr
 &&+H_{F1} \Delta_{\sigma\sigma}\,\delta F_{tz})\Big),
 \eea
where, for $\{\xi,\eta\}$ taking values in the set $\{\sigma,t\}$, we have
\bea
\gamma_\xi &=& \gamma_1 \frac{\partial \phi_1}{\partial \xi} +\gamma_2 \frac{\partial \phi_2}{\partial \xi},  \\[2mm] 
\Delta_{\xi \eta} &=& \frac{\partial r}{\partial \xi}\frac{\partial r}{\partial \eta} + r^2\frac{\partial \theta }{\partial \xi}\frac{\partial \theta}{\partial\eta} 
+ r^2 \sin^2 \theta \frac{\partial \phi_1}{\partial \xi}\frac{\partial \phi_1}{\partial \eta} + r^2 \cos^2 \theta\frac{\partial \phi_2}{\partial \xi} \frac{\partial \phi_2}{\partial\eta}, \quad {\rm and} \\
\quad \psi_{t\sigma} &=& \frac{\partial \phi_1}{\partial t}\frac{\partial \phi_2}{\partial \sigma}-\frac{\partial \phi_1}{\partial \sigma}\frac{\partial \phi_2}{\partial t}.
\eea
After explicitly implementing the embedding via 
\be\label{emb}\{r=r(t),\theta=\theta(t), \phi_1=\sigma,\phi_2=\phi_2(t)\}, \ee 
these reduce to 
\bea
\Delta_{\sigma\sigma} = r^2\sin^2\theta, \quad \Delta_{\sigma t} &= &0, \quad \Delta_{tt}= \dot{r}^2+r^2\dot{\theta}^2 + r^2\cos^2\theta\,\dot{\phi_2}^2, \\
\gamma_{\sigma} = \gamma_1,  \quad \gamma_t &=& \gamma_2\,\dot{\phi_2}, \quad \psi_{t\sigma}=-\dot{\phi_2},
\eea
 and $\L$ reduces to (\ref{arg6}).

In \cite{virmani} the Lagrangian was first computed without reference to a specific embedding.  However, we have found that due to the gauge ambiguity, the proper WZ term to use depends on which embedding is chosen.  Namely, it turns out that equations (\ref{c3}) and (\ref{c7}) for $C^{(3)}$ and $C^{(7)}$ are valid for the embedding with $\phi_2=\sigma$, but if instead $\phi_1=\sigma$ (which of course is the focus of most of the paper), $C^{(3)}$ and $C^{(7)}$ are given by (\ref{c3sin}) and (\ref{c7sin}).  For the latter case we arrive at the WZ term given by 
\bea
(\tau_{D6} \,k)^{-1}\L_{WZ} &=& 
\Big(H_{D4}^{-1}(1+\gamma_2\,\dot{\phi_2})-1+  B_0^2\,[H_{D0}^{-1}(1+\gamma_2\,\dot{\phi_2})-1]\Big) F_{z \sigma}  \cr
&& +(B_0^2 H_{D0}^{-1}+H_{D4}^{-1})\gamma_1\,\delta F_{tz} +  (Q_{D0}+B_0^2 Q_{D4})\, \dot{\phi_2}\sin^2\theta.  \label{wz6}
\eea  
We also note that expressing $\delta F_{tz}$ in terms of conserved quantities gives
\be\label{deltaftz}
\delta F_{tz}=\frac { F_{z \sigma}^2 \, r^2[j_1/\tau_{D6} V_6- (H_{D0} +  B_0^2 H_{D4})r^2\sin^2\theta] }
{r^2 (H_{D0} +  B_0^2 H_{D4})(F_{z \sigma}^2 H_{F1} r^2 + H_{D0} H_{D4} r^4\sin^2\theta -2\omega F_{z \sigma}\sin^2\theta)\sin^2\theta },
\ee
and we can do the same for the $\phi_2$ component of the linear velocity:
\be\label{vphi2}
v_{\phi_2}= |r\cos\theta\dot{\phi_2}| = \left \lvert \frac{F_{z \sigma} r^3 [j_2/\tau_{D6} V_6- (Q_{D0} +  B_0^2 Q_{D4})\sin^2\theta] }{r^2 (H_{D0} +  B_0^2 H_{D4})(F_{z \sigma}^2 H_{F1} r^2 + H_{D0} H_{D4} r^4\sin^2\theta -2\omega F_{z \sigma}\sin^2\theta)\cos\theta} \right\rvert.
\ee 

Lastly, we can also mention the other worldvolume supertube construction in \cite{bena}, the ``superposition" supertube, in which the three-charge supertube is realized as a superposition of two-charge supertubes.  Since this object is constructed from worldvolumes of different species of D-brane it is not a proper supertube, however.  Specifically, it is created from a superposition of a set of $k_2$ D2-brane supertubes with D0 and F1 charge, superimposed upon $k_6$ D6-brane supertubes of the same shape and radius with D4 and F1 charge.  It is also possible to add a collection of $k_5$ NS5-brane supertubes with D0 and D4 charge.  Naturally, the D6-branes and NS5-branes are wrapped on the $T^4$.  We note that the superposition option is not available for the supergravity supertubes because nonlinear interactions come into play \cite{elvang}.    

\section{The Embedding Radius}  \label{appb}

The worldvolume interval is given by
\be
ds^2=g_{ab}dx^a dx^b= g_{tt}dt^2 + 2g_{t\sigma}dt \,d\sigma + 2g_{tz}dt\, dz + g_{\sigma\sigma}d\sigma^2 + 2g_{z \sigma}d\sigma\,dz + g_{zz}dz^2 + ds^2_{T^4}\, ,
\ee
where the worldvolume metric $g_{ab}$ is the pullback of the spacetime metric $G_{\mu\nu}$ to the worldvolume of the supertube: 
\be g_{ab} = G_{\mu\nu}\frac{\partial x^{\mu}}{\partial x^{a}}\frac{\partial x^{\nu}}{\partial x^{b}}\, .
\ee 
With the embedding of Section \ref{embed} the components $g_{ab}$ are 
\bea
g_{tt} &=& 
-H_{D0}^{-1/2}H_{D4}^{-1/2}H_{F1}^{-1} (1 + \frac{\omega}{r^2}\cos^2\theta\,\dot{\phi_2})^2
 + H_{D0}^{1/2}H_{D4}^{1/2} (\dot{r}^2+r^2\dot{\theta}^2+r^2\cos^2\theta\dot{\phi}_2^2 ),\nonumber\\[2mm]
g_{t\sigma} &=&  
- H_{D0}^{-1/2} H_{D4}^{-1/2} H_{F1}^{-1}( \frac{\omega}{r^2}\sin^2\theta  +\frac{\omega^2}{r^4}\sin^2\theta\cos^2\theta\,\dot{\phi_2}), \nonumber\\[2mm]
 g_{\sigma\sigma} & = & 
  H_{D0}^{1/2} H_{D4}^{1/2}r^2\sin^2\theta- H_{D0}^{-1/2} H_{D4}^{-1/2}H_{F1}^{-1} \,\frac{\omega^2}{r^4}\sin^4\theta\, ,  \nonumber\\[2mm]
g_{zz} &=& H_{D0}^{1/2} H_{D4}^{1/2} H_{F1}^{-1}\, , \nonumber\\[2mm]
g_{tz} &=& g_{z \sigma}=0,  \nonumber\\
g_{66} &= &g_{77}=g_{88}=g_{99}=H_{D0}^{1/2} H_{D4}^{-1/2},
\eea
keeping in mind that the metric is in the string frame and $\omega$ is the angular momentum parameter of the black hole from (\ref{bhspin}).  In the BPS limit, the velocities $\{\dot{r}, r\,\dot{\theta}, r\cos\theta\dot{\phi}_2\}$ vanish.  Furthermore, we can change bases by switching from $\{dt,d\sigma\}$ to $\{e^0, d\tilde{\sigma}\}$ where $e^0=dt+\frac{\omega}{r^2}\sin^2\theta d\sigma$ and $d\tilde{\sigma}=d\sigma$, obtaining 
\bea
ds^2 &=& -H_{D0}^{-1/2} H_{D4}^{-1/2} H_{F1}^{-1}(e^0)^2 + H_{D0}^{1/2} H_{D4}^{1/2}r^2\sin^2\theta d\tilde{\sigma}^2 + H_{D0}^{1/2} H_{D4}^{1/2} H_{F1}^{-1}dz^2 + ds^2_{T^4}  \nonumber \\
&\equiv& -H_{D0}^{-1/2} H_{D4}^{-1/2} H_{F1}^{-1}(e^0)^2 + R^2 d\tilde{\sigma}^2 + H_{D0}^{1/2} H_{D4}^{1/2} H_{F1}^{-1}dz^2 + ds^2_{T^4}. 
\eea
We now have a worldvolume interval that is free of off-diagonal terms and more suited to a notion of proper circumference.  What we call the embedding radius $R=R(\vec{X})$ is the proper circumference of the supertube divided by $2\pi$, that is $R^2=g_{\tilde{\sigma}\tilde{\sigma}}$.  Explicitly,  
\be
R^2= H_{D0}^{1/2}H_{D4}^{1/2}r^2 \sin^2\theta=(Q_{D0}+r^2)^{1/2}(Q_{D4}+r^2)^{1/2}\sin^2\theta.  
\ee
It is perhaps useful to point out that the transformation $dt\rightarrow dt+\frac{\omega}{r^2}\sin^2\theta d\sigma$ 
does not lead to a globally defined timelike coordinate due to the fact that the $S^1_{\sigma}$ is noncontractible.\footnote{We thank D. Marolf for correspondence on this point.}  
This prevents us from introducing a coordinate $\tilde{t}$ such that $d\tilde{t}=e^0$.

The corresponding Type IIB solution is obtained by T-dualizing the spacetime IIA fields on the $z$ coordinate (see e.g. \cite{virmani}); the resulting pulled-back metric has nonzero $g_{t\sigma}$, $g_{tz}$ and $g_{z \sigma}$.  In this case we switch from $\{dt,d\sigma, dz\}$ to $\{e^0, d\tilde{\sigma}, e^5\}$ where 
\be e^0=dt+\frac{\omega}{r^2}\sin^2\theta d\sigma, \quad e^5=dz+\frac{\omega}{r^2}\sin^2\theta d\sigma, \quad \mbox{and} \quad d\tilde{\sigma}=d\sigma. \ee
As before, $R^2$ is just $g_{\tilde{\sigma}\tilde{\sigma}}$, which gives
\be R^2= (Q_{D1}+r^2)^{1/2}(Q_{D5}+r^2)^{1/2}\sin^2\theta \ee
for charge parameters $Q_{D1}$ and $Q_{D5}$.  A property of the T-dual metric is that $Q_{D1}=Q_{D0}$ and $Q_{D5}=Q_{D4}$.  Thus we see that the embedding radius for the IIB supertube is the same as that of the IIA one. 

\section{Supersymmetric Black Rings: Physical Parameters}\label{appc}

With black rings, the compilation of solutions is not as complete as for black holes.  Supersymmetric black rings were first discovered in \cite{mateos2}; three-charge, three-dipole versions (and their string theoretic descriptions) were given in \cite{elvang,warner2,gauntlett2} and the general family of their non-supersymmetric deformations is believed to have nine parameters.  To date, however, only a smaller seven-parameter family of these non-supersymmetric solutions is known \cite{elvang2}.  The BPS black rings themselves are characterized by up to seven independent parameters. One possible choice of these seven would be the charges (which are conserved), the dipoles (which are not conserved), and $\R$, which some refer to as the radius of the ring.  Some expressions for the BPS physical quantities are given below.\footnote{The actual result of a fragmentation merger would be non-BPS, but the BPS quantities are close approximations for the near-extremal solutions in which we are interested.}  The integer-valued charges and dipoles we label as 
\be 
\quad\{N_1, N_2, N_3\} =\{N_{D0}, N_{F1}, N_{D4}\}, \quad \mbox{and} \quad \{n^1, n^2, n^3\} = 
\{n_{D6}, n_{NS5}, n_{D2}\}. 
\ee
Following the notation of \cite{virmani}, the ADM mass and angular momenta are given by
\bea
M &=& 
\frac{1}{ g_s\sqrt{\alpha'} } \left(N_{D0}+ \frac{ \ell^4}{\alpha'^2}N_{D4} + \frac{g_s R_z}{\sqrt{\alpha'}}N_{F1}\right),\\
J_2 &=& \frac{1}{2} \left( n_{D6} N_{D0} + n_{D2} N_{D4} + n_{NS5}N_{F1} - n_{D6}\, n_{D2}\,
n_{NS5} \right), \\
J_1 &=& J_2 + \frac{\pi \R^2}{4 G_5} \left( g_s \sqrt{\alpha'}n_{D6}  +  \frac{g_s \alpha'^{5/2}}{\l^4} n_{D2} + \frac{\alpha'}{R_z} n_{NS5} \right),
\eea
while the entropy is 
\be \label{Sring} S = 2 \pi \sqrt{N_{D0} N_{D4} N_{F1} - \N_{D0}\N_{D4}\N_{F1} - J_2^2 - n_{D6}\, n_{D2}\, n_{NS5} (J_1 -J_2)},
\ee
where 
\be \N_{D0} = N_{D0} - n_{D2}\, n_{NS5}, \quad \N_{D4} = N_{D4} - n_{D6}\, n_{NS5}, \quad
\N_{F1} = N_{F1} - n_{D2}\, n_{D6}.
\ee

\end{document}